\documentclass[aps,prl,preprint,superscriptaddress]{revtex4-2}

\usepackage{graphicx} 
\usepackage{subfigure}
\usepackage{amsmath}
\usepackage{float}
\usepackage[T1]{fontenc}
\usepackage[applemac]{inputenc}
 \usepackage{xcolor}

\begin{document}

\title{Time-Reversal Mirror inside a granular suspension: \\ a way of measuring the ultrasound diffusion coefficient}

\author{Yamil Abraham}

\affiliation{Institut Langevin, ESPCI Paris, Universit\'e PSL, CNRS, Paris, France}

\affiliation{Instituto de F\'isica, Facultad de Ciencias, Universidad de la Rep\'ublica, Montevideo, Uruguay}

\author{Bart A. van Tiggelen}

\affiliation{Laboratoire de Physique et Mod\'elisation des Milieux Condens\'es, Universit\'e Grenoble Alpes, CNRS, Grenoble, France}

\author{Nicolas Benech}

\affiliation{Instituto de F\'isica, Facultad de Ciencias, Universidad de la Rep\'ublica, Montevideo, Uruguay}

\author{Carlos Negreira}

\affiliation{Instituto de F\'isica, Facultad de Ciencias, Universidad de la Rep\'ublica, Montevideo, Uruguay}
 
\author{Xiaoping Jia}

\affiliation{Institut Langevin, ESPCI Paris, Universit\'e PSL, CNRS, Paris, France}

\author{Arnaud Tourin}
\email[]{arnaud.tourin@espci.psl.eu}

\affiliation{Institut Langevin, ESPCI Paris, Universit\'e PSL, CNRS, Paris, France}

\date{\today}

\begin{abstract}
We demonstrate that the diffusion coefficient, $D$, for ultrasound propagating in a multiple scattering medium, such as a dense granular suspension, can be measured using a time reversal experiment. This requires an unprecedented experimental setup in which a piezoelectric transducer, acting as a Time-Reversal Mirror (TRM), is embedded within the granular suspension at a depth much larger than the scattering mean free path, while an array of transducers is placed in the far field of the scattering sample. A single element of the array emits a short pulse and the TRM records the resulting ultrasonic field, which consists of a ballistic coherent wave followed by a coda wave. When the entire coda wave is time-reversed and re-emitted from the TRM, it is observed to refocus on the original source, with a focal spot size that decreases with the inverse depth of the TRM, characteristic of a diffusive regime. Time-reversal of short windows selected at different times $t$ in the coda wave reveals a focal spot size that decreases as the inverse square root of time, i.e., $1 / \sqrt{Dt}$. By fitting the predictions of a microscopic diffusion theory to our experimental data, we are able to accurately measure the diffusion coefficient in the granular suspension. Remarkably, this method does not require ensemble averaging due to stability of time-reversal against statistical fluctuations. 

\end{abstract}


\maketitle


Multiple scattering is a regime in which a wave propagating through a random collection of obstacles is scattered more than once. When the scattering medium becomes much larger than the scattering mean free path, the transport of energy is well described by the classical diffusion equation \cite{sheng2006introduction}, as successfully tested by Page et al. for ultrasonic waves propagating in a concentrated suspension of glass beads immersed in water \cite{page1995experimental}. Unlike light, the use of ultrasound has the distinct advantage that the time-dependent wave-field can be easily detected. By combining time-resolved amplitude and intensity measurements, in both transmission and reflection, all transport parameters (scattering, transport and absorption mean free paths, diffusion coefficient) can be determined, as described in \cite{tourin2000transport}.

Interestingly, while the diffusion equation describes intrinsically irreversible processes at the macroscopic level, the wave equation keeps its time-reversal invariance at the microscopic level (before averaging over disorder). Pioneering experiments were carried out in the mid-1990s to test the reversibility of a multiply scattered ultrasonic wave, whose average energy density is described by the diffusion equation. To this end, Derode and al. \cite{derode1995robust} employed a Time Reversal Mirror (TRM), i.e., a device capable of capturing a broadband wavefield and re-emitting it in reverse chronological order, resulting in a wave that converges back onto its source. Their experiments led to a rather counterintuitive result:  not only is it possible to make a wave revive its past in a strongly scattering medium, but the wave is actually focused better than it would be in a homogeneous medium. As demonstrated in \cite{tourin1999dynamic}, the scattering medium can be regarded as a random lens that allows a broad angular spectrum of the source to be collected by a small-aperture TRM, even if the latter is limited to a single transducer. Furthermore, the more disordered the lens, the more efficient the spectrum recovery \cite{tourin2006phononic}. This has contributed to creating a new paradigm for the manipulation of waves in complex media: contrary to long-held beliefs, disorder is not merely an obstacle to focusing and imaging but can be harnessed as an ally to control waves.

Here we advance this concept by demonstrating that a time-reversal experiment can even be used to measure the ultrasound diffusion coefficient. To achieve this, we adopt the configuration proposed by Van Tiggelen \cite{van2003green} where the TRM is buried deep within the scattering sample. We show that this unprecedented experimental setup creates a situation where the effective aperture of the random lens is solely governed by diffusion, and the size of the focal spot therefore varies as the inverse of the diffusive halo size at the sample exit.

\begin{figure}[h!]
	\centering
	\includegraphics[width=.45 \textwidth]{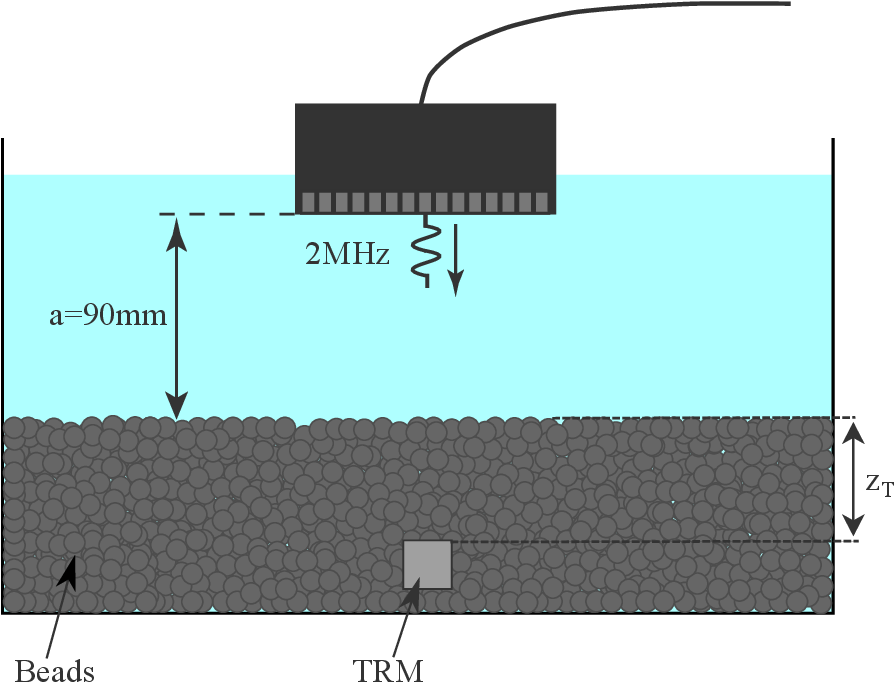}
	\caption{Schematic representation of the experimental setup.}
	\label{fig:montaje_generico_TRyTP}
\end{figure}

\textit{Experiment} - The scattering sample we consider consists of a random close-packed suspension of monodisperse glass beads, each 1.5mm in diameter (see figure \ref{fig:montaje_generico_TRyTP}). This sample is immersed in a water tank measuring 400 mm in length, 150 mm in width and 200 mm in height. The beads are placed at the bottom of the tank and carefully mixed before each measurement in order to create a new realization of disorder, achieving each time a volume fraction close to the random close-packing limit, i.e., $\phi \simeq 0.64$. An ultrasonic array of piezoelectric transducers, with a central frequency of 2MHz, is also immersed in the water tank at a distance of $a=90$mm from the sample surface. The transducer array consists of 64 rectangular elements, each 0.315mm wide ($\sim$ half-wavelength in water at 2MHz) and 15mm high, with a pitch of 0.375mm. A single-element transducer, consisting of a rectangular piezoelectric element, 0.50mm wide and 12mm high, with a central frequency of 1.5MHz, serves as the TRM. This transducer is held at the bottom of the water tank at a depth $z_T$ from the sample surface using a dedicated support. The length $z_T$ is chosen to be much larger than the scattering mean free path, which is of the order of one bead diameter \cite{schriemer1997energy}. This choice ensures a diffusive regime, where a source located at depth $z_T$ creates a diffusive halo with a size $\sim z_T$ at the sample exit. 

\begin{figure}[h!]
	\centering
	\subfigure{\includegraphics[width=.45 \textwidth]{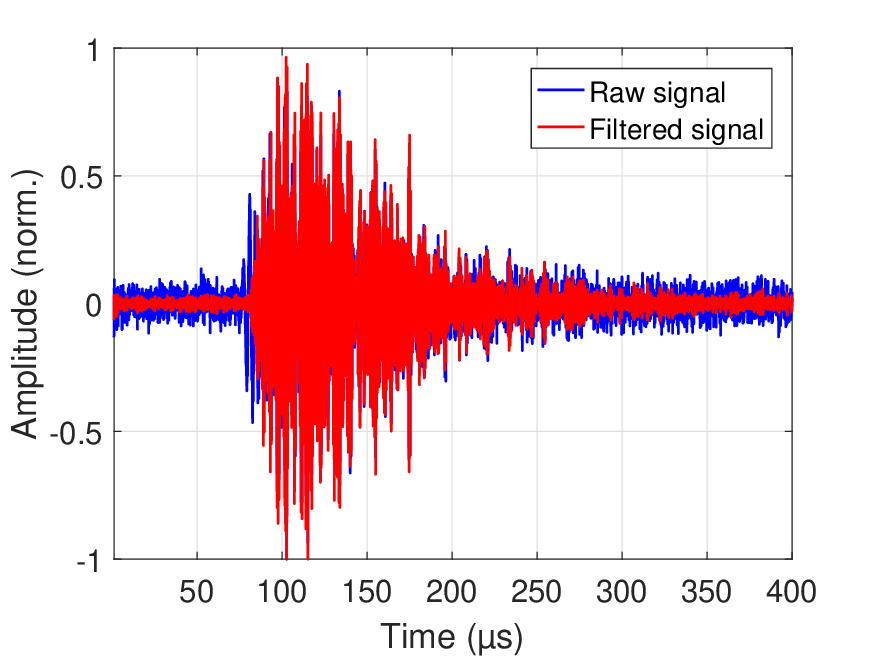}}
	\subfigure{\includegraphics[width=.45 \textwidth]{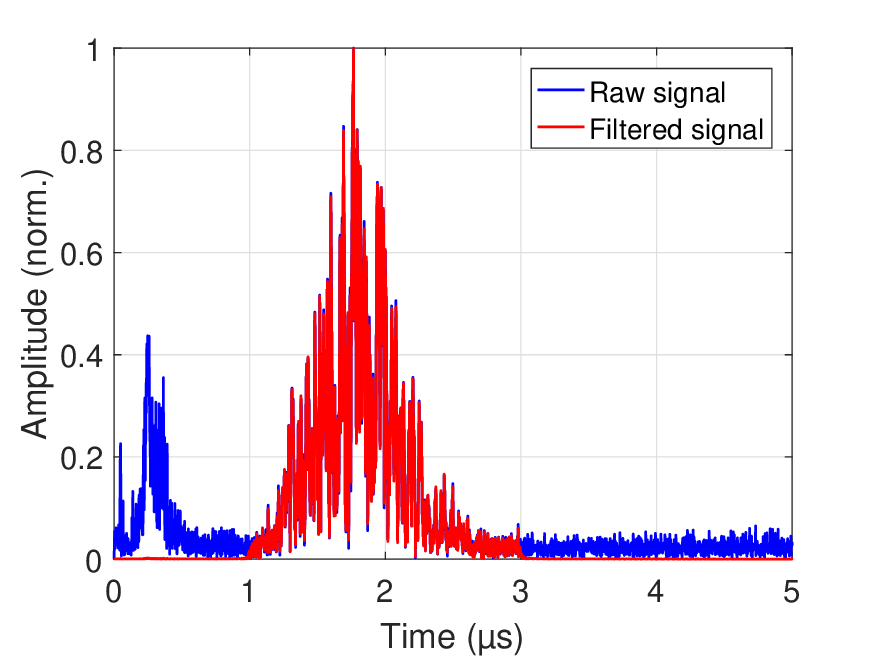}}
	\caption{(a) A typical waveform recorded on the TRM (in blue) is composed of a first low-frequency arrival, i.e, the ballistic coherent wave, followed by a long scattering signal usually called the coda wave. It is frequency-filtered to suppress the ballistic wave, resulting in the signal shown in red, before time-reversal. (b) Spectra of the recorded waveform before (blue) and after filtering (red)}
	\label{fig:ballistic_coda}
\end{figure}

A time-reversal experiment is carried out in two steps. First, one element in the array is used as a source to emit a 2-cycle sinusoidal wave at a central frequency of 2MHz. The resulting transmitted waveform is then measured by the TRM. As seen in figure \ref{fig:ballistic_coda}, this waveform consists of a low-frequency ballistic coherent wave (i.e., the part of the wave that would resist averaging over disorder) followed by a long scattering signal, commonly referred to as the coda wave.
A  bandpass filter, ranging from 1 to 3MHz, is applied to eliminate the low-frequency ballistic contribution, as well as to reduce the electronic noise. In the second step, a time window, called the Time Reversal Window (TRW), is selected from the transmitted signal and re-emitted. The TRW can encompass the entire recorded signal (\textit{stationary time-reversal experiment}), or a smaller part of it (\textit{dynamic time-reversal experiment}.). 

\begin{figure}[h!]
\centering
\subfigure[$t_{w}=93\mu s$]{\includegraphics[width=.400 \textwidth]{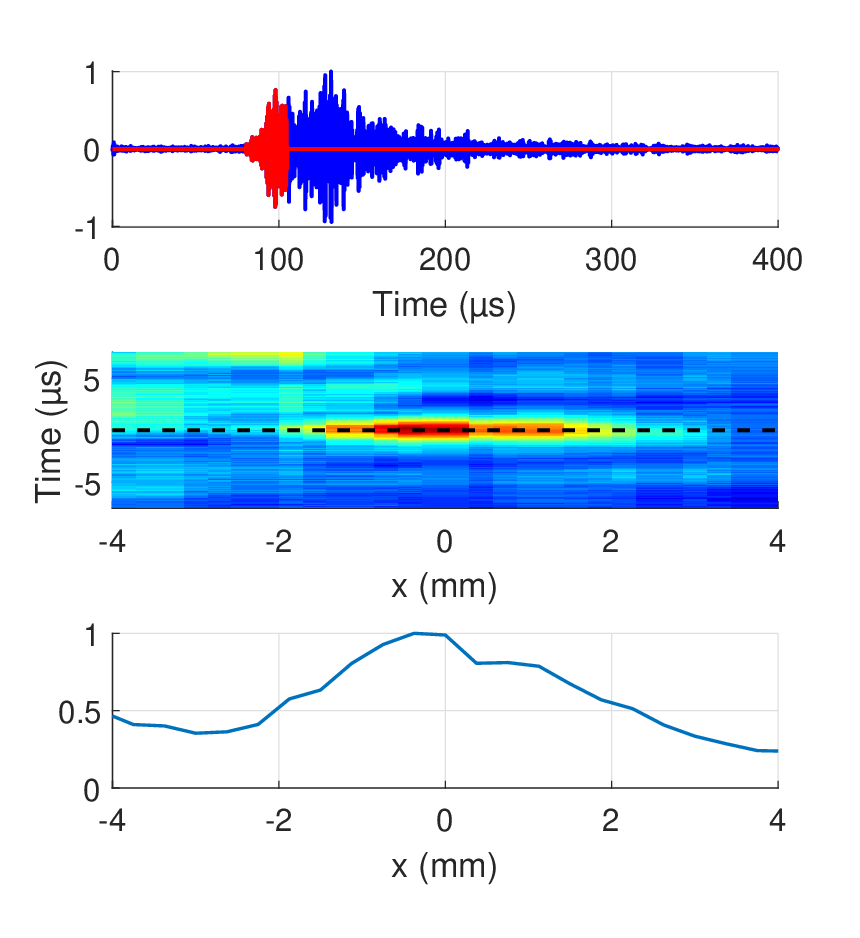}}
\subfigure[$t_{w}=133\mu s$]{\includegraphics[width=.400 \textwidth]{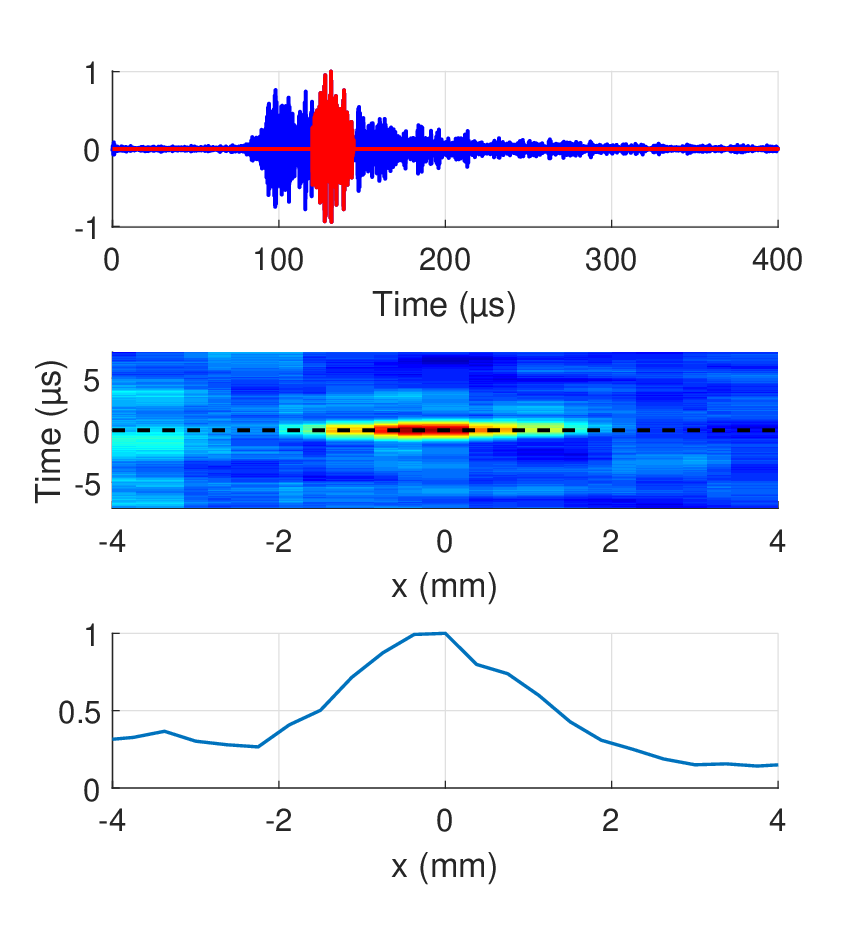}}

\subfigure[$t_{w}=148\mu s$]{\includegraphics[width=.400 \textwidth]{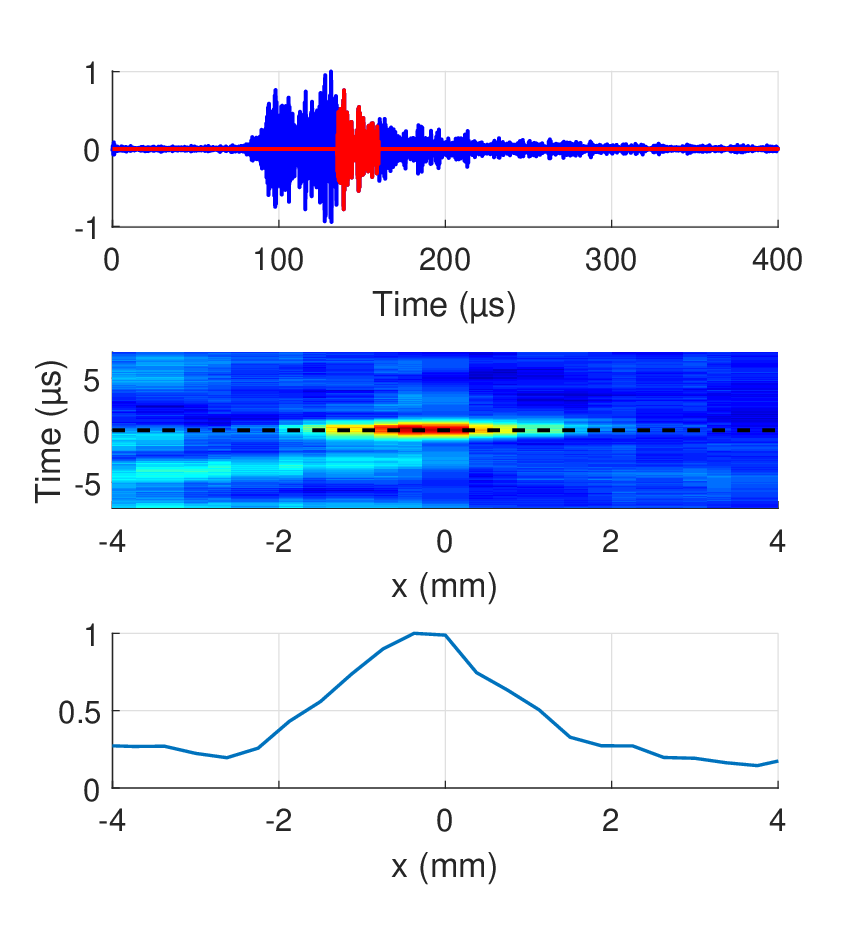}}
\subfigure[$t_{w}=193\mu s$]{\includegraphics[width=.400 \textwidth]{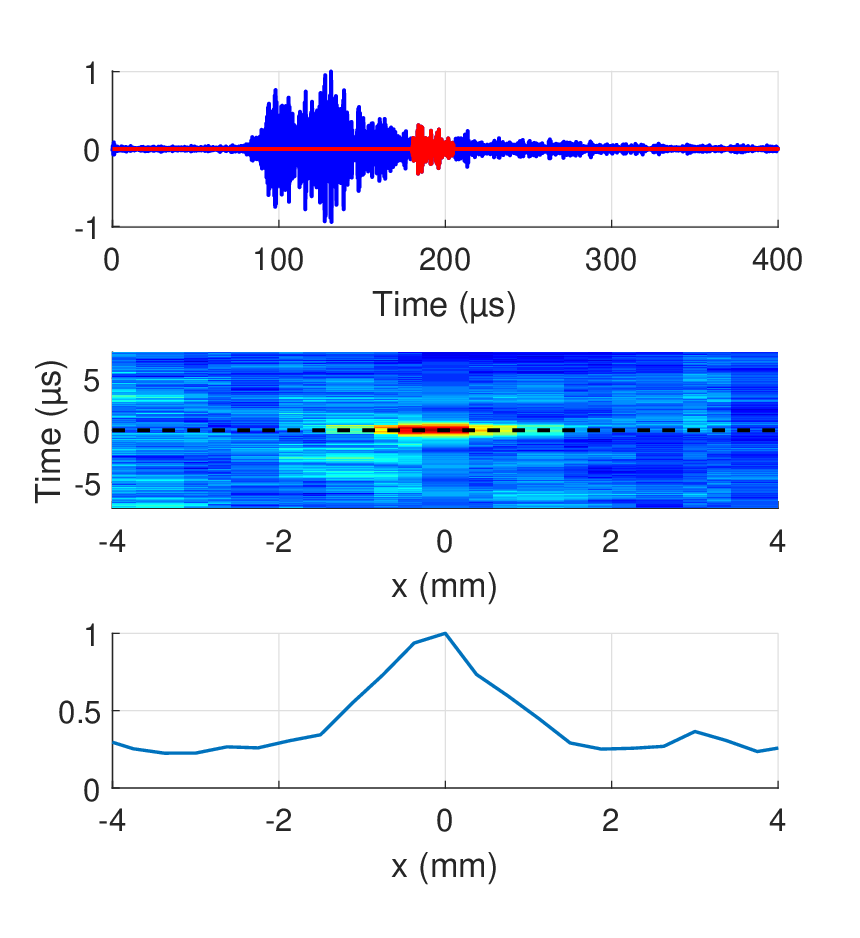}}
	\caption{Dynamic time-reversal for four different Time Reversal Windows (TRW). Top views: a TRW (in red) is selected in the recorded signal and time-reversed. Middle views: the resulting time-reversed signal measured at the array is focused in both time and space. The signal at the source recovers its initial duration (inverse of the bandwidth). Bottom views: cross sections at the moment of focusing, defined as $t=0$. The focal spot size decreases as a function of the TRW, as can be seen from figures (a) to (d).}
	\label{fig:focus_vs_t_beads15mm}
\end{figure}

For \textit{dynamic time-reversal experiments}, we set  $z_T$ to 25mm. We select a sliding TRW of length of $T=25\mu s$, starting at time $t_{w}-T/2$ (shown in red in the top views of figures \ref{fig:focus_vs_t_beads15mm}.a,b,c,d). A time step of $5\mu$s is chosen between consecutive TRWs, resulting in a $20\mu$s overlap. After backpropagation, the resulting time-reversed wave amplitude is then recorded by all elements of the ultrasonic array and plotted as a function of space and time (see the middle views in figures \ref{fig:focus_vs_t_beads15mm}.a,b,c,d). The focal spot is determined from the amplitude at each transducer at the moment of focusing (bottom views in figures \ref{fig:focus_vs_t_beads15mm}.a,b,c,d). It is important to note that the chosen TRW size results from a compromise. On the one hand, it should not be too short because, for both time-reversed peak and spatial focusing, we expect the signal-to-noise ratio (defined as the peak amplitude divided by the standard deviation of the surrounding noise) to scale with the square root of the ratio of TRW size to the initial pulse duration \cite{derode2001random}, (i.e., 5 for our choice of a $25-\mu s$ long window). On the other hand, the TRW should not be too long, so as to remain sensitive to the temporal evolution of the focal spot. The four examples shown in figure \ref{fig:focus_vs_t_beads15mm} illustrate how the time-reversed focal spot size  decreases as the TRW is selected further into the coda wave for a given realization of disorder. 
Finally, the  \textit{dynamic-time reversal experiment} is repeated for a few other disorder configurations to check the statistical robustness of the measurement. By stirring the glass bead packing inside the water tank each time, we have access to 8 different configurations. 

\begin{figure}
	\centering 
	\includegraphics[width=.45 \textwidth]{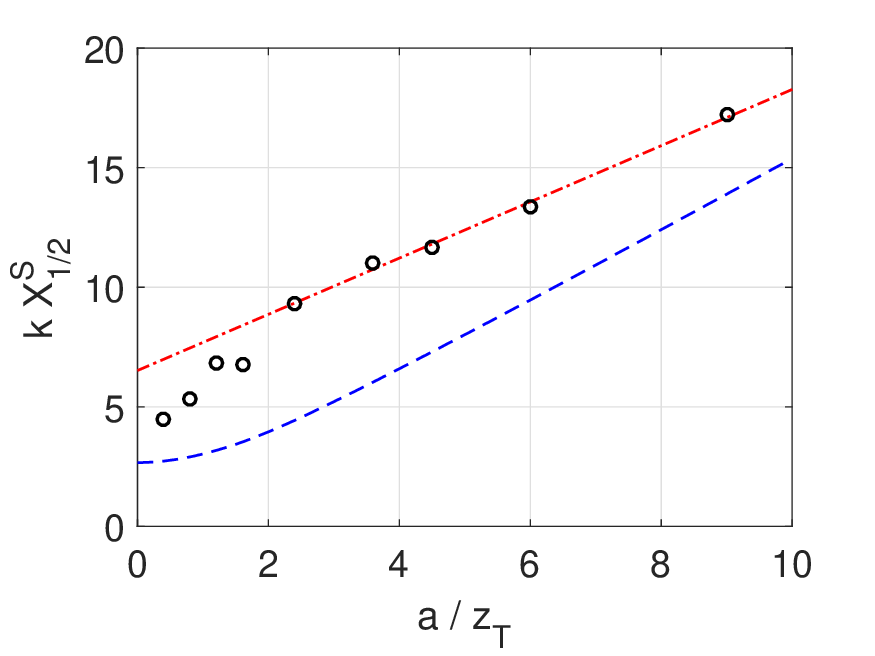}
		\caption{Evolution of focal spot size as a function of TRM depth in a stationary time reversal experiment. The red dashed line is a linear fit to the data for $a / z_T > 2$ and has slope 1.3. The theoretical prediction (blue line with a slope 1.7) takes into account the directivity of the source as well as the finite TRW size (see figure \ref{fig:Stationary_TR_Depth} in the Supplemental Material for the influence of these parameters on the predicted focal spot size).}
	\label{fig:delta_vs_zT_15mm}
\end{figure}

For the \textit{stationary time-reversal experiment}, we need to select the longest possible TRW, at least greater than the Thouless time $z_T^2/2D$ \cite{van2003green}. We thus choose a TRW of size 400 $\mu$s (limited by the noise level in the recorded signal), which is of the order of the Thouless time for the largest TRM depth ($z_T=25 mm$). We observe that the resulting focal spot size varies inversely as the TRM depth (see figure \ref{fig:delta_vs_zT_15mm}), which is a signature of an isotropic diffusion regime as discussed below.

\begin{figure}[h!]
	\centering
	\subfigure[]{\includegraphics[width=.300 \textwidth]{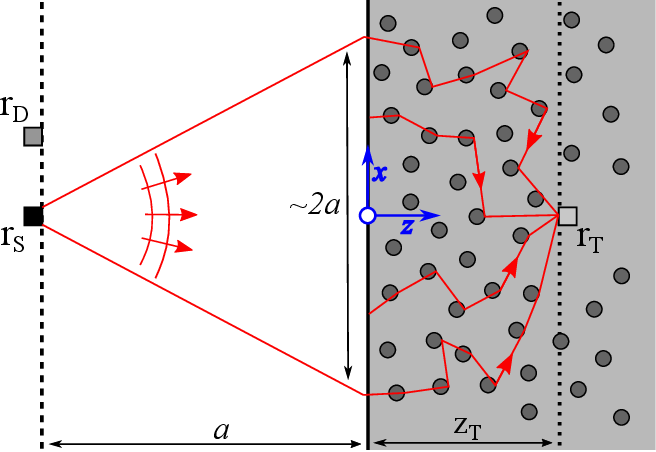}}
	\hspace{20mm}
	\subfigure[]{\includegraphics[width=.300 \textwidth]{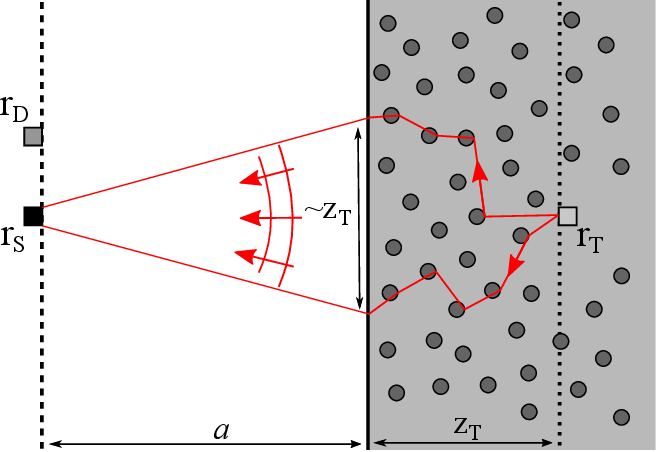}}
	\subfigure[]{\includegraphics[width=.300 \textwidth]{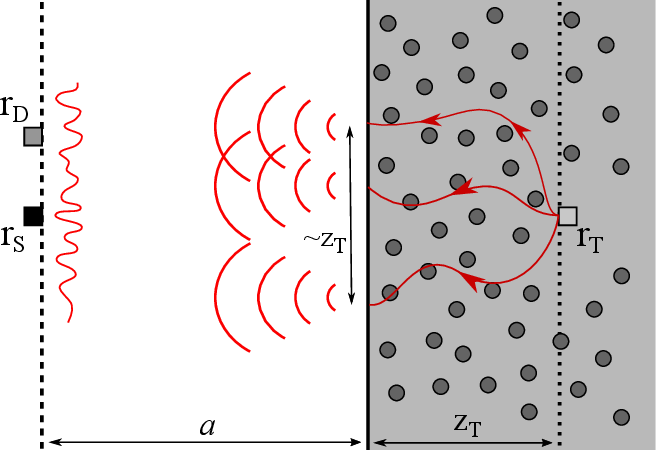}}
	\caption{Schematic diagram of the experimental set-up and coordinate system used for microscopic scattering theory.  For clarity, we use dark gray disks to represent scattering events. (a) A half-wavelength-sized source (one of the elements of the transducer array) emits a wavefront that reaches the scattering sample as a cylindrical beam with a transverse size $2a$ along $x$. Due to scattering, the angular spectrum of the source is redirected towards the TRM, albeit limited to a single transducer. In this sense, the sample acts as a random lens. (b) When the waveform measured at $\mathbf{r}_T$ is time-reversed and sent back, the focal spot size of the refocused signal at $\textbf{r}_S$ is determined by the transverse size of the wavefront emerging from the sample, which in the best case is given by $2a$. (c) The focal spot size can also be viewed as an estimator of the spatial correlation function of the scattered field measured at distance $a$ from the sample exit for a source located at $z_T$. In the isotropic diffusion regime, the halo at the sample exit has a of size $z_T$ and consists of random, uncorrelated sources. Consequently, the Van Cittert-Zernike theorem applies, and the correlation length is given by $a/ kz_T$ provided that $ a \gg z_T$.}
	\label{fig:esquema_montaje_TRyTP}
\end{figure}

\textit{Theory } - This result, as well as the observed decrease of the focal spot size in the dynamic experiment, is well-described by the microscopic diffusion theory we have developed (see Supplemental Material for the description of the full theory), the main points of which are outlined here. We assume the multiple scattering medium to be semi-infinite in the half space $z>0$ (see figure \ref{fig:esquema_montaje_TRyTP}.a). 
When a Dirac pulse is emitted from the point-like source at $\mathbf{r}_S$ (one element in the array), the signal received at the TRM at $\mathbf{r}_T$ is $G\left( \mathbf{r}_T , \mathbf{r}_S , t\right)$; i.e., the Green's function of the wave equation between points $\mathbf{r}_S$ and $\mathbf{r}_T$. A TRW is then selected to produce $G\left( \mathbf{r}_T , \mathbf{r}_S , t\right) rect\left( \frac{t - t_w}{T}\right)$, where $rect$ is the rectangle function, producing a time window of size $T= 25 \mu s$ around time $t_w$. The corresponding signal is time-reversed and sent back from $\mathbf{r}_T$ to the array. The signal received at a position $\mathbf{r}_D$ at the array is then given by

\bigskip

\begin{equation*}
	\begin{split}
		F_{t_w} \left(|\mathbf{r}_D-\mathbf{r}_S|, t \right) =
		\left[ rect \left( \frac{- t - t_w}{T}\right) G \left(\mathbf{r}_T , \mathbf{r}_S, -t \right) \right]
		\otimes G\left(\mathbf{r}_D , \mathbf{r}_T , t \right)
		\\
		= \int d\tau  G \left(\mathbf{r}_T , \mathbf{r}_S , \tau \right) 
		G \left( \mathbf{r}_D , \mathbf{r}_T , t + \tau \right) rect \left( \frac{\tau-t_w}{ T} \right)
		\\
		= \int_{t_w - T/2} ^ {t_w + T/2} d\tau  G \left( \mathbf{r}_T , \mathbf{r}_S , \tau \right)
		G \left(\mathbf{r}_D , \mathbf{r}_T , t + \tau \right) 
	\end{split}
\end{equation*}
\bigskip

Considering $G (\mathbf{r}_T , \mathbf{r}_S ,t)=G(\mathbf{r}_S , \mathbf{r}_T ,t)$ due to reciprocity, the focal spot size can be determined by analyzing, at the focusing time $t=0$, the variation of 

\begin{equation}
 F_{t_w}(X)=F_{t_w} (|\mathbf{r}_D-\mathbf{r}_S| , 0) = \int_{t_w - T/2} ^ {t_w + T/2} d \tau G \left( \mathbf{r}_S , \mathbf{r}_T , \tau \right) G(\mathbf{r}_D , \mathbf{r}_T , \tau)
\label{eq:corr} 
\end{equation}
as a function of  the distance $X= \| \mathbf{r}_D -\mathbf{r}_S \|$ around the initial source. This expression is a statistical estimator of the spatial correlation function $\langle G \left( \mathbf{r}_S , \mathbf{r}_T , t_w \right) G \left( \mathbf{r}_D , \mathbf{r}_T , t_w \right) \rangle$ that would be measured between points $\mathbf{r}_S$ and $\mathbf{r}_D$ for a point source located at $\mathbf{r}_T$ \cite{derode2001random}.

In the stationary case (which involves integrating Eq. \ref{eq:corr} over an infinite time), the spatial correlation function can be predicted as $C(X)$=exp$(- z_T kX/a)$ with $k$ the wavenumber in the background medium (here water). This prediction is based on Akkermans and Maynard's formula for the Coherent Backscattering (CBS) \cite{akkermans1986coherent} generalized in \cite{van2003green} for a TRM placed inside a scattering medium (see figure \ref{fig:Stationary_TR_Depth} in the Supplemental Material). The characteristic size of C(X), given by $a /k  z_T$, indicates that the transverse size of the diffusive halo at the sample exit should correspond to the depth of the source within the scattering sample (see figure \ref{fig:esquema_montaje_TRyTP}.b). The celebrated Van-Cittert Zernike theorem \cite{goodman2015statistical} indeed states that $a /k z_T$  is the correlation length of a wave field generated by a fully spatially incoherent random source of size $z_T$ at a distance $a$. In other words, in a \textit{stationary time reversal}, the focal spot size should vary inversely with the depth of the TRM. This is what we observe as soon as $a/ z_T \gg 1 $ (see figure \ref{fig:delta_vs_zT_15mm}), although the slope differs from the predicted $ln(2)$ as derived from $C(X)$=exp$(- z_T kX/a)$. Actually, the $ln(2)$ slope would imply an infinite TRW as suggested by the microscopic diffusion theory we developed. When accounting for the finite size of the TRW (see blue curve in figure \ref{fig:delta_vs_zT_15mm} obtained for a TRW of $400 \mu s$ as in our stationary time reversal experiment), the agreement between experiment and theory improves. However, it remains imperfect, as the theory does not account for all experimental parameters, particularly the finite aperture of the TRM. Note that for $ a / z_T < 1$, this linear dependence of the focal spot size on inverse depth breaks down due the failure of the far-field assumption. Furthermore, the y-intercept is non zero since the focal spot size cannot  be smaller than the initial source size. And for a source size smaller than the wavelength, the focal spot size cannot be smaller than half the wavelength (diffraction limit) of ultrasound near the transducer array (i.e., in water). 

For the focal spot obtained in a \textit{dynamic time-reversal experiment}, the full calculation (see Supplemental Material) gives: 
\begin{equation}
\begin{split}
	F_{t} \left( X \right) \propto \int_{0}^{\theta_{max}} d\theta  \sin \theta
	J_0(kX \sin \theta) e^{-a^2 \tan^2 \theta /4Dt}
	\times \left[ \frac{e^{-z_T^2/4Dt}-e^{-\left( z_T+2z_0 \right)^2/4Dt} }{ \left( Dt \right)^{3/2}} \right]
	\label{eq:tr_dynamic}
	\end{split}
\end{equation}
with $\theta_{max}$ fixed to $\pi/4$ by the finite directivity of the source used in our experiment.

\bigskip

From Eq. \eqref {eq:tr_dynamic}, we deduce that the focal spot size varies as $ a/ k\sqrt{Dt}$ (see figure \ref{fig:Dynamic_TR}  in the Supplemental Material for the evolution of the focal spot size against time), which indicates that the diffusion process creates a halo of size $\sqrt{Dt}$ at the sample exit. This result is reminiscent of the Coherent Backscattering (CBS) \cite{Tsang:84, wolf1985weak, van1985observation, bayer1993weak}, especially in its dynamic version \cite{tourin1997time}, which is not surprising given the close link between Time Reversal and CBS \cite{van2003green, de2004relation}. More quantitively, it comes out from the theory that the half-width at half-maximum of the focal spot is given by:
\begin{equation}
\Delta = \frac{0.82 a}{k \sqrt{Dt}}. 
\label{eq:prediction}
\end{equation}

\begin{figure}
	\centering 
	\subfigure[]{\includegraphics[width=.45 \textwidth]{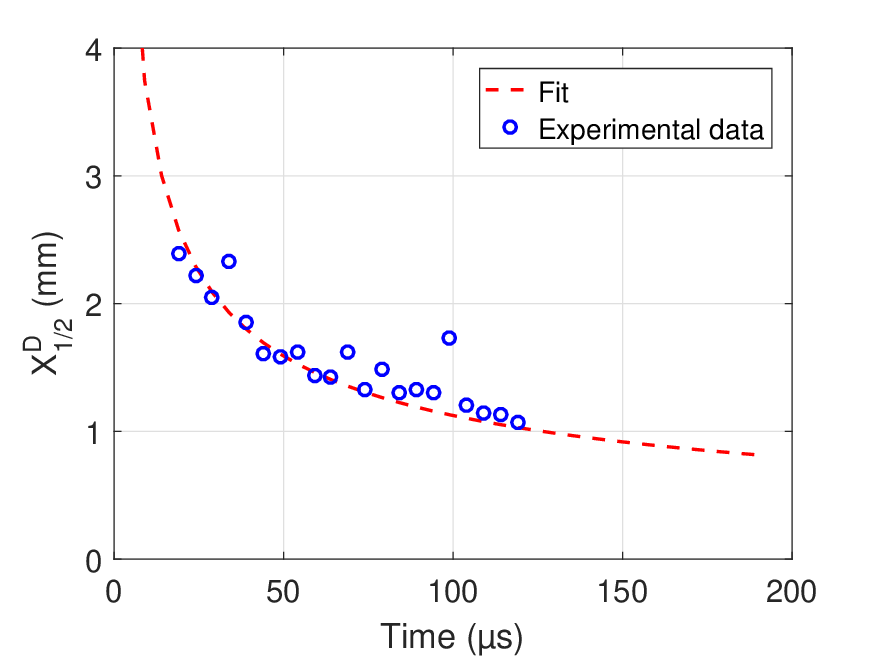} \label{fig:delta_d_a}}
	\subfigure[]{\includegraphics[width=.45 \textwidth]{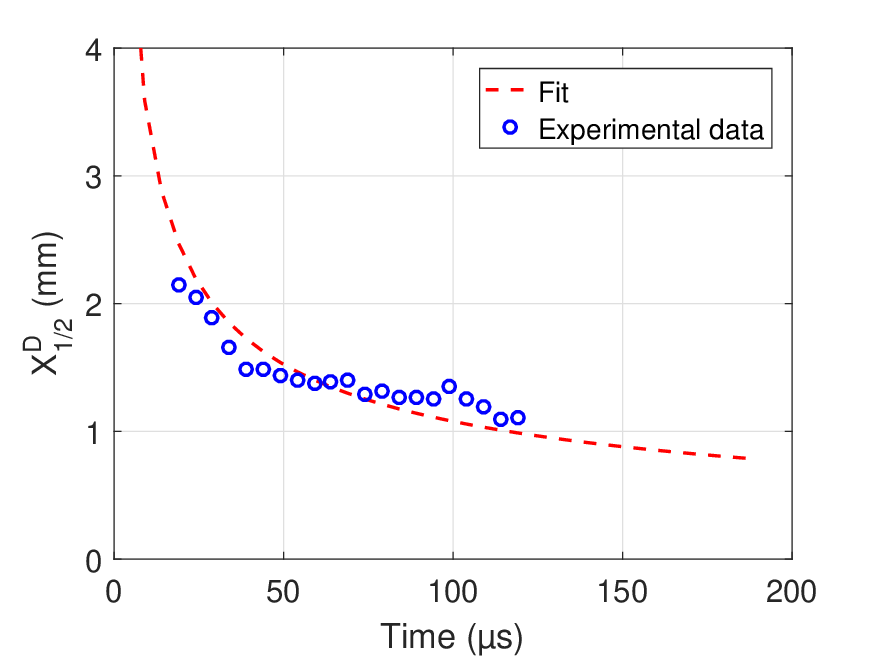} \label{fig:delta_d_b}}
	\caption{Half-width at half-maximum of the focal spot as a function of the TRW position. The time axis has been shifted to align the origin with the activation time of the diffusive sources, accounting for the travel time in water. The experimental results (blue circles) are fitted using equation \eqref{eq:prediction} (red dashed line). (a) Single realization of disorder. (b) Focal spots are determined after averaging the time-reversed wave field amplitudes over 8 realizations.}
	\label{fig:delta_d}
\end{figure}

\textit{Discussion} - Figure \ref{fig:delta_d}.a shows the evolution of the experimental focal spot size against time for a given realization of disorder. Fitting this data to Eq. \eqref {eq:prediction} yields $D=0.61 \pm 0.08$ $\mathrm{mm^2 \mu s^ {-1}}$, where $0.08 \mathrm{mm^2 \mu s^ {-1}}$ represents the fit uncertainty.  Remarkably, this measurement is obtained without ensemble-averaging, unlike conventional methods (either exploiting the time-dependent average transmitted intensity or the CBS) that require averaging over disorder. Here, we leverage the stability of Time Reversal against statistical fluctuations \cite{Papanicolaou2002}. Time Reversal is in fact known to be a self-averaging process, as the average over disorder is replaced by an average over the number of uncorrelated pieces of information measured by the TRM, called spatial and/or temporal \textit{information grains} in Ref. \cite{derode2001random}. In our case, we have only one spatial information grain (the single-channel TRM) but we have $N =T /  \tau = \triangle \omega / \delta \omega =5 $ temporal/frequency information grains in the TRW (with $\tau$ the initial pulse duration, inverse of the bandwidth $\triangle\omega$ and $\delta \omega$ the decorrelation frequency). Provided that $D$ does not vary significantly over the bandwidth, the measurement method proposed here is thus stable against statistical fluctuations. To fully assess the statistical robustness of the method, we repeated the \textit{dynamic time-reversal experiment} for 8 realizations of disorder and found values slightly scattered around the mean value $D=0.65 \pm 0.14 $ $\mathrm{mm^2 \mu s^ {-1}}$. Notably, the spread around the average value is of the same order as the fitting uncertainty for each disorder realization, confirming the statistical robustness of the method. Figure \ref{fig:delta_d}.b finally shows the time evolution of the focal spots obtained after ensemble averaging the time-reversed waves over the 8 realizations. Fitting this data to Eq. \eqref {eq:prediction} yields a value of $D=0.67 \pm 0.07 $ $\mathrm{mm^2 \mu s^ {-1}}$, consistent with the previous measurement.

These values are in close agreement with those deduced, using the same set-up, from the two conventional methods mentioned above. On the one hand, we used the transducer array to measure the dynamic CBS (for which the full 3D theory can be found in \cite{cobus2017dynamic}). This method yielded $D=0.70 \pm 0.03$ $\mathrm{mm^2 \mu s^ {-1}}$. On the other hand, we brought the transducer array into contact with the scattering sample and used one of its elements as a source and the TRM as a receiver to measure the time dependence of the transmitted intensity. Its average over 8 realizations was then fitted to the solution of the diffusion equation giving $D=0.74 \pm 0.02$ $\mathrm{mm^2 \mu s^ {-1}}$. It should be pointed out that these values are not averaged over exactly the same bandwidth. In the second case, as in our time-reversal experiments, the central frequency of the measured transmitted wave field is slightly downshifted to a lower frequency (1.75 MHz) compared to the central frequency of the emitted pulse (see figure \ref{fig:ballistic_coda}). This frequency shift occurs because the central frequency of the TRM is 1.5 MHz. In contrast, when recording the CBS, the transducer array is used for both emitting and receiving, meaning that D is averaged over the array bandwidth, which is centered at 2MHz. 

\textit{Conclusion} - In this paper, we have reported on ultrasound time-reversal experiments performed with a TRM buried deep inside a scattering sample, specifically a random suspension of monodisperse glass beads. In this unprecedented experimental configuration, the scattering medium creates an effective aperture that is governed solely by diffusion. 
In a \textit{stationary time-reversal experiment}, this effective aperture is given by the TRM depth so that the focal spot size scales inversely with depth, as we have verified experimentally. This behavior can be understood through the following physical reasoning: the focal spot can be viewed as an estimator of the spatial correlation function of the speckle pattern that would be measured on the transducer array for a source located at depth $z_T$ within the medium. In a regime of isotropic diffusion, the diffusive halo at the sample exit has a transverse size $z_T$ and consists of random, uncorrelated sources. Consequently, the Van Cittert-Zernike theorem applies, and the correlation length is thus given by $a/ kz_T$ provided that $a \gg z_T$. 

Beyond its fundamental significance, this result is of practical importance, as it reveals the possibility of determining the depth of an active source embedded in a cluttered environment by analyzing the spatial correlation of the scattered field it generates in the far field of the region of interest. This a topic of broad relevance across various fields, with examples in optical fluorescence-based imaging of biological media  \cite{PhysRevLett.121.023904}, ultrasonic non-destructive testing \cite{PhysRevLett.98.104301}, and electromagnetic source localization \cite{Scientific_Report}.

In a \textit{dynamic time-reversal experiment}, we find that the focal spot size varies as $1/\sqrt Dt$ where $t$ is the TRW position, i.e., the inverse size of the diffusive halo at the sample exit. This approach provides a novel method for measuring the diffusion coefficient, which, unlike conventional techniques, does not require ensemble averaging due to the inherent stability of TR against statistical fluctuations.


\begin{acknowledgments}
\textit{Acknowledgments} - This work has received support under the program "Investissements d'Avenir" launched by the French Government. This work was supported by Agencia Nacional de Investigaci\'on e Innovaci\'on (ANII), Uruguay. We acknowledge Maxime Harazi who has conducted the very first experiments during his PhD thesis. 
\end{acknowledgments}

\bibliography{bibliography.bib}


\pagebreak

\setcounter{equation}{0}
\setcounter{figure}{0}
\setcounter{page}{1}
\renewcommand{\theequation}{S\arabic{equation}}
\renewcommand{\thefigure}{S\arabic{figure}}
\renewcommand{\bibnumfmt}[1]{[S#1]}
\renewcommand{\citenumfont}[1]{S#1}

\begin{center}
  \textbf{\large Supplemental Material 
\\Time-Reversal Mirror inside a granular suspension: \\ a way of measuring the ultrasound diffusion coefficient}\\[.2cm]
 \end{center}
 
\begin{figure}[H]
\begin{center}
\includegraphics[height=7cm]{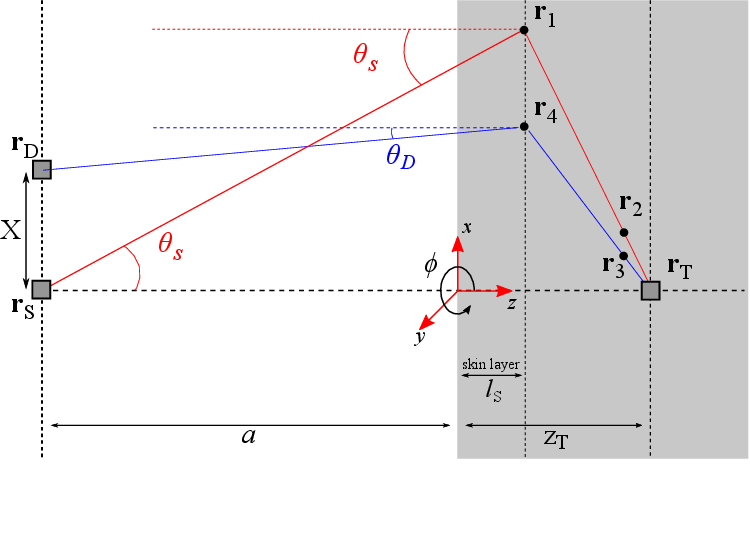}
\caption{Notations for the theoretical development.}
\label{fig:Notacion_Cuentas_BvT_2}
\end{center}
\end{figure}

The multiple scattering medium is assumed to be semi-infinite in the half space $z>0$ (see figure \ref{fig:Notacion_Cuentas_BvT_2}).

When a Dirac pulse is emitted from the point source at $\textbf{r}_S$, the signal received at $\textbf{r}_T$ is 
 $G(\textbf{r}_T,\textbf{r}_S,t)$; i.e., the Green's function between points $\textbf{r}_S$ and $\textbf{r}_T$. A part of this signal, referred to as the Time-Reversal Window (TRW), is then selected to produce  $G (\textbf{r}_T,\textbf{r}_S,t )W( \frac{t-t_w}{ T})$ where $W$ is the rectangle function. The corresponding signal is  time reversed and sent back. Finally the signal received at position $\textbf{r}_D$ is given by

\begin{equation*}
\begin{split}
F_{t_w}(\textbf{r}_D,\textbf{r}_S,t)=\left[W( \frac{-t-t_w}{ T})G (\textbf{r}_T,\textbf{r}_S,-t )\right] \bigotimes G(\textbf{r}_D,\textbf{r}_T,t)
\\
=\int W( \frac{\tau-t_w}{ T}) G(\textbf{r}_T,\textbf{r}_S,\tau) G(\textbf{r}_D,\textbf{r}_T,t+\tau) d \tau
\\
=\int_{t_w-T/2}^{t_w+T/2} G(\textbf{r}_T,\textbf{r}_S,\tau) G(\textbf{r}_D,\textbf{r}_T,t+\tau) d \tau
\end{split}
\end{equation*}

\bigskip

Under the assumption of reciprocity, $G(\textbf{r}_T,\textbf{r}_S,t)=G(\textbf{r}_S,\textbf{r}_T,t)$ so that $F_{t_w}(\textbf{r}_D,\textbf{r}_S,t)$ can be rewritten as 

\begin{equation*}
F_{t_w}(\textbf{r}_D,\textbf{r}_S,t)
=\int_{t_w-T/2}^{t_w+T/2} G(\textbf{r}_S,\textbf{r}_T,\tau) G(\textbf{r}_D,\textbf{r}_T,t+\tau) d \tau
\end{equation*}

\bigskip

The width of the focal spot can thus be determined by analyzing at time $t=0$ the variation of  $F_{t_w}(\textbf{r}_D,\textbf{r}_S,t=0)=\int_{t_w-T/2}^{t_w+T/2} G(\textbf{r}_S,\textbf{r}_T,\tau) G(\textbf{r}_D,\textbf{r}_T,\tau) d \tau $  as a function of the distance $X=\|\textbf{r}_D -\textbf{r}_S \|$ around the source, which is a statistical estimator of the spatial correlation function $\langle G(\textbf{r}_S,\textbf{r}_T, t_w) G(\textbf{r}_D,\textbf{r}_T,t_w) \rangle$ for a point source located in $\textbf{r}_T$ \cite{derode2001random}. The directivity pattern obtained by time-reversing a window positioned at $t$ will thus be predicted as $d^D_{t}(X)=\langle G(\textbf{r}_S,\textbf{r}_T, t) G(\textbf{r}_D,\textbf{r}_T,t) \rangle$. This spatial correlation function can be expressed using a traditional transport operator, as we shall do in the following. The calculation is more tractable in the frequency domain, which asks for a Laplace Transform: 

\begin{equation*}
C(\textbf{r}_S,\textbf{r}_D,\Omega)
=\int_{0}^{\infty} \langle G(\textbf{r}_S,\textbf{r}_T, t) G(\textbf{r}_D,\textbf{r}_T,t)\rangle e^{i(\Omega t +i \epsilon)} dt
\end{equation*}

\bigskip

By decomposing $G(\textbf{r}_S,\textbf{r}_T, t)$ and $G(\textbf{r}_D,\textbf{r}_T,t)$ on their frequency components, we can write:
\begin{equation*}
\begin{split}
C(\textbf{r}_S,\textbf{r}_D,\Omega)=\int_{0}^{\infty} \int_{-\infty}^{\infty} \int_{\infty}^{\infty} d \omega_1 d \omega_2 dt \langle G(\textbf{r}_S,\textbf{r}_T, \omega_1) G(\textbf{r}_D,\textbf{r}_T, \omega_2) \rangle e^{-i\omega_1 t} e^{-i\omega_2 t} e^{i(\Omega + i \epsilon) t }
\end{split}
\end{equation*}

\bigskip

Exploiting the fact that $\int_{0}^{\infty} e^{-i(\Omega + i \epsilon- \omega_1 - \omega_2)t} dt=\delta(\Omega + i \epsilon- \omega_1 - \omega_2)$, we get
\begin{equation*}
\begin{split}
C(\textbf{r}_S,\textbf{r}_D,\Omega)= \int_{-\infty}^{\infty} \langle G(\textbf{r}_S,\textbf{r}_T, \omega_1) G(\textbf{r}_D,\textbf{r}_T, \Omega + i \epsilon - \omega_1)\rangle d \omega_1
\end{split}
\end{equation*}

$G(\textbf{r}_D,\textbf{r}_T,t)$ is a real function so that $G(\textbf{r}_D,\textbf{r}_T, \omega)=G(\textbf{r}_D,\textbf{r}_T,-\omega )^*$. 

\bigskip

By using the change of variables $\omega_1 \rightarrow \omega + \Omega / 2 +i \epsilon /2$, we obtain:

\begin{equation*}
\begin{split}
C(\textbf{r}_S,\textbf{r}_D,\Omega)= \int_{-\infty}^{\infty} \langle G(\textbf{r}_S,\textbf{r}_T, \omega+\Omega / 2 + i \epsilon /2) G^*(\textbf{r}_D,\textbf{r}_T, \omega-\Omega / 2 - i \epsilon /2)\rangle d \omega
\end{split}
\end{equation*}

Here we use the fact that $G(\textbf{r}_S,\textbf{r}_T, \omega)$ and $G(\textbf{r}_D,\textbf{r}_T, \omega)$ can be analytically continued in the lower sheet Im$\omega > 0$ since $G(\textbf{r}_D,\textbf{r}_T, t<0)=0$  and $G(\textbf{r}_S,\textbf{r}_T, t<0)=0$. The frequency function under the integral sign determines the time aspects of the propagation.

\bigskip

The object that determines the exact microscopic space-time behavior of a disturbance is thus $\langle G(\textbf{r}_S,\textbf{r}_T, \omega+\Omega / 2 +i\epsilon) G^*(\textbf{r}_D,\textbf{r}_T, \omega-\Omega / 2-i\epsilon)\rangle$, which obeys the following equation \cite{LAGENDIJK1996143}:

\begin{multline}
\label{TR1}
\langle G(\textbf{r}_S,\textbf{r}_T, \omega+\Omega / 2+i\epsilon) G^*(\textbf{r}_D,\textbf{r}_T, \omega-\Omega / 2-i\epsilon)\rangle
\\
=
G_e(\textbf{r}_S,\textbf{r}_T, \omega+\Omega / 2+i\epsilon) G_e^*(\textbf{r}_D,\textbf{r}_T, \omega-\Omega / 2-i \epsilon)
\\
+ \int_{1} \int_{2} \int_{3} \int_{4} d\textbf{r}_1 d\textbf{r}_2 d\textbf{r}_3 d\textbf{r}_4
G_e(\textbf{r}_S,\textbf{r}_1, \omega+\Omega / 2+i\epsilon) G_e^*(\textbf{r}_D,\textbf{r}_4, \omega-\Omega / 2-i\epsilon) \Gamma (\textbf{r}_1, \textbf{r}_2, \textbf{r}_3, \textbf{r}_4)
\\
 G_e(\textbf{r}_2,\textbf{r}_T, \omega+\Omega / 2+i\epsilon) G_e^*(\textbf{r}_3,\textbf{r}_T, \omega-\Omega / 2-i\epsilon) \end{multline}
where $G_e$  is the effective (average) Green's function and $\Gamma$ is the Vertex.

\bigskip

At this step, we make the following hypothesis:

\begin{enumerate}

\item

The TRM is supposed to be at a depth $z_T \gg l_s$ such that the first term on the right hand-side can be disregarded.

\item

The frequency $\omega$ of the internal oscillations is much larger than the frequency of the enveloppe $\Omega$. 

\item

In the ladder approximation, the Vertex writes  $\Gamma (\textbf{r}_1, \textbf{r}_2, \textbf{r}_3, \textbf{r}_4)=\tilde{P}(\textbf{r}_1, \textbf{r}_2)\delta(\textbf{r}_2-\textbf{r}_3)\delta(\textbf{r}_1-\textbf{r}_4)$ where $\tilde{P}(\textbf{r}_1, \textbf{r}_2)$ describes transport from $\textbf{r}_1$ to $\textbf{r}_2$.

\end{enumerate}

\bigskip

 Eq. \eqref{TR1} thus simplifies to:
\begin{multline*}
\label{TR1}
\langle G(\textbf{r}_S,\textbf{r}_T, \omega+\Omega / 2+i\epsilon) G^*(\textbf{r}_D,\textbf{r}_T, \omega-\Omega / 2-i\epsilon)\rangle=
\\
 \int_{1} \int_{2}  G_e(\textbf{r}_S,\textbf{r}_1, \omega+i\epsilon) G_e^*(\textbf{r}_D,\textbf{r}_1, \omega-i\epsilon) \tilde{P}(\textbf{r}_1, \textbf{r}_2) G_e(\textbf{r}_2,\textbf{r}_T, \omega+i\epsilon) G_e^*(\textbf{r}_2,\textbf{r}_T, \omega-i\epsilon) d\textbf{r}_1 d\textbf{r}_2
\end{multline*}
where only the propagator $\tilde{P}$ still depends on $\Omega$ on the right-hand side.

\bigskip

After an inverse Laplace Transform over $\Omega$ , we obtain
\begin{multline*}
d^D_t(X)=\langle G(\textbf{r}_S,\textbf{r}_T, t) G(\textbf{r}_D,\textbf{r}_T, t)\rangle=
\\
 \int_{1} \int_{2}  G_e(\textbf{r}_S,\textbf{r}_1, \omega) G_e^*(\textbf{r}_D,\textbf{r}_1, \omega) P(\textbf{r}_1, \textbf{r}_2,t) G_e(\textbf{r}_2,\textbf{r}_T, \omega) G_e^*(\textbf{r}_2,\textbf{r}_T, \omega) d\textbf{r}_1 d\textbf{r}_2
 \end{multline*}
where $t$ is the position of the TRW measured with respect to the activation time of the diffusive sources at $z=l^*$, where $l^*$ is the transport mean free path (assumed to be equal to the scattering mean free path $l_s$).

Assuming no internal reflection, we can use the theorem of images with boundary condition $P=0$ at the trapping plane $z=-z_0$ (with $z_0=\frac{2}{3}  l^*$ in 3D  \cite{lagendijk1989influence}) to deduce from the solution of the diffusion equation in an infinite medium the propagator $P(\textbf{r}_1, \textbf{r}_2,t)$ linking the points $\textbf{r}_1$ and $\textbf{r}_2$ in half-space $z>0$.
\begin{equation*}
P(x_1,x_2, y_1,y_2, z_1,z_2,t)=\frac{e^{-\rho_{12}^2/4Dt}}{(Dt)^{3/2}}(e^{-(z_1-z_2)^2/4Dt}-e^{-(z_1+z_2+2z_0)^2/4Dt})
\end{equation*}
where $\rho_{12}=\sqrt{(x_1-x_2)^2+(y_1-y_2)^2}$ is the lateral distance between points $\textbf{r}_1$ and $\textbf{r}_2$ along the sample interface.

\bigskip

And we have:

\begin{equation*}
|G_e(\textbf{r}_2,\textbf{r}_T, \omega)|^2=\frac{ e^{-|\textbf{r}_2 - \textbf{r}_T | / \ell_S}}{(4\pi)^2|\textbf{r}_2 - \textbf{r}_T |^2} 
\end{equation*}

\begin{equation*}
G_e(\textbf{r}_S,\textbf{r}_1, \omega)=\frac{e^{ik |\textbf{r}_S - \textbf{r}_1 |}}{4\pi|\textbf{r}_S - \textbf{r}_1 |} e^{-\frac{z_1} {2 \ell_S \cos \theta_S}}
 \end{equation*}

\begin{equation*}
G^*_e(\textbf{r}_D,\textbf{r}_1, \omega)=\frac{e^{-ik |\textbf{r}_D - \textbf{r}_1 |}}{4\pi|\textbf{r}_D- \textbf{r}_1 |} e^{-\frac{z_1}{2 \ell_S \cos \theta_D}}
\end{equation*}

\bigskip
\bigskip
In the last two Green's functions, we neglect the phase accumulated in the skin layer (see figure \ref{fig:Notacion_Cuentas_BvT_2}). Furthermore, if $|\textbf{r}_S - \textbf{r}_D |<<a $, one can assume $\cos \theta_S=\cos \theta_D=\mu $

\bigskip

The integral we have to perform is thus:
\begin{multline*}
\langle G(\textbf{r}_S,\textbf{r}_T, t) G(\textbf{r}_D,\textbf{r}_T, t)\rangle
\\ 
\\ 
= \int...\int dx_1 dx_2  dy_1 dy_2 dz_1 dz_2\frac{1}{(4\pi)^4}   \frac{ e^{-\frac{|\textbf{r}_2 - \textbf{r}_T |}{\ell_S}}}{|\textbf{r}_2 - \textbf{r}_T |^2}  \frac{e^{ik |\textbf{r}_S - \textbf{r}_1 |}}{|\textbf{r}_S - \textbf{r}_1 |} \frac{e^{-ik |\textbf{r}_D - \textbf{r}_1 |}}{|\textbf{r}_D - \textbf{r}_1 |} e^{-z_1 / \mu \ell_S} P(x_1,x_2, y_1,y_2, z_1,z_2,t)
\\
\\ 
\simeq \int \int \int dz d^2\rho \frac{e^{ik |\textbf{r}_S - \textbf{r} |}}{|\textbf{r}_S - \textbf{r} |} \frac{e^{-ik |\textbf{r}_D - \textbf{r} |}}{|\textbf{r}_D - \textbf{r} |} e^{-z/ \mu \ell_S} \frac{e^{-\rho^2/4Dt}}{(Dt)^{3/2}}[e^{-(z-z_T)^2/4Dt}-e^{(z+z_T+2z_0)^2/4Dt}]  
\end{multline*}
 \\ 
since $ e^{-\frac{|\textbf{r}_2 - \textbf{r}_T |}{\ell_S}}/|\textbf{r}_2 - \textbf{r}_T |^2 \sim \delta(\textbf{r}_2 - \textbf{r}_T)$.
 \\ 
 \\ 
 We have for any point $\textbf{r}$ at the sample surface
 
 \begin{equation*}
 |\textbf{r}_S - \textbf{r}|=\sqrt{a^2+\rho^2}
\end{equation*}

\begin{equation*}
\begin{split}
 |\textbf{r}_D - \textbf{r}| & =\sqrt{a^2+(\rho \cos \phi-X)^2 +(\rho \sin \phi)^2} \\
  &=\sqrt{a^2+\rho^2 + X^2-2\rho X \cos \phi)}\\
 & = \sqrt{a^2+\rho^2}\sqrt{  1+\frac{X^2-2\rho X \cos \phi}{a^2+\rho^2}  }\\
& \simeq \sqrt{a^2+\rho^2} [1-(\rho X \cos \phi)/(a^2+\rho^2)]\\
\end{split}
\end{equation*}

Thus,
\begin{multline*}
\langle G(\textbf{r}_S,\textbf{r}_T, t) G(\textbf{r}_D,\textbf{r}_T, t)\rangle
\\ 
\simeq \int_0^\infty \rho d\rho \int_0^{2\pi} d\phi  \int_0^\infty dz  \frac{e^{-z / \mu \ell_S }}{a^2+\rho^2}  e^{ik \frac{\rho X \cos \phi}{\sqrt{a^2+\rho^2}}} \frac{e^{-\rho^2/4Dt}}{(Dt)^{3/2}}[e^{-(z-z_T)^2/4Dt}-e^{-(z+z_T+2z_0)^2/4Dt}]
\end{multline*}

 \bigskip

Since the term depending on $z$ under the integral does not vary much over a thickness $\ell_S$, one can write  $\int_0^\infty dz e^{-z / \mu \ell_S }f(z)\simeq \mu \ell_S f(0)$.
\bigskip

We thus have:
\begin{multline*}
\langle G(\textbf{r}_S,\textbf{r}_T, t) G(\textbf{r}_D,\textbf{r}_T, t)\rangle
\simeq \int_0^\infty \rho d\rho \int_0^{2\pi} d\phi \frac{\mu}{a^2+\rho^2}  e^{ik \frac{\rho X \cos \phi}{\sqrt{a^2+\rho^2}}} \frac{e^{-\rho^2/4Dt}}{(Dt)^{3/2}}[e^{-z_T^2/4Dt}-e^{-(z_T+2z_0)^2/4Dt}]
\end{multline*}

\bigskip

We then use $\cos \theta = a / \sqrt{a^2 + \rho^2}$, $\sin \theta = \rho / \sqrt{a^2 + \rho^2}$, $\tan \theta = \rho/a$ et $d\rho= a/\cos^2 \theta d\theta$ so that
\begin{multline*}
\langle G(\textbf{r}_S,\textbf{r}_T, t) G(\textbf{r}_D,\textbf{r}_T, t)\rangle
\\
 \simeq \int_{0}^{\theta_{max}} d\theta  \sin \theta \frac{e^{-a^2 \tan^2 \theta /4Dt}}{(Dt)^{3/2}}[e^{-z_T^2/4Dt}-e^{-(z_T+2z_0)^2/4Dt}] \int_0^{2\pi} d\phi  e^{ik(sin \theta)X \cos \phi}
 \\
 \simeq \int_{0}^{\theta_{max}} d\theta \sin \theta  \frac{e^{-a^2 \tan^2 \theta /4Dt}}{(Dt)^{3/2}}[e^{-z_T^2/4Dt}-e^{-(z_T+2z_0)^2/4Dt}] J_0(kX \sin \theta)
\\ 
\end{multline*}
Finally, we end up with the following formula:

\begin{equation}
 \begin{split}   
d^D_{t}(X)&=\langle G(\textbf{r}_S,\textbf{r}_T, t) G(\textbf{r}_D,\textbf{r}_T, t)\rangle
\\
&\simeq   \int_{0}^{\theta_{max}} d\theta \sin \theta e^{-a^2 \tan^2 \theta /4Dt} J_0(kX \sin \theta) \frac{[e^{-z_T^2/4Dt}-e^{-(z_T + 2 z_0)^2/4Dt}]}{(Dt)^{3/2}}
 \end{split} 
\label{eq:dyn} 
\end{equation}

It should be pointed out that the value of $\theta_{max}$ depends on the source directivity. For a point-like source, $\theta_{max}= \pi /2$. In our experiment, $\theta_{max}$ is limited to $\pi /4$ due to the source finite size ($\sim \lambda/2$) 

\begin{figure}
\centering 
 \subfigure[]{\includegraphics[width=.45 \textwidth]{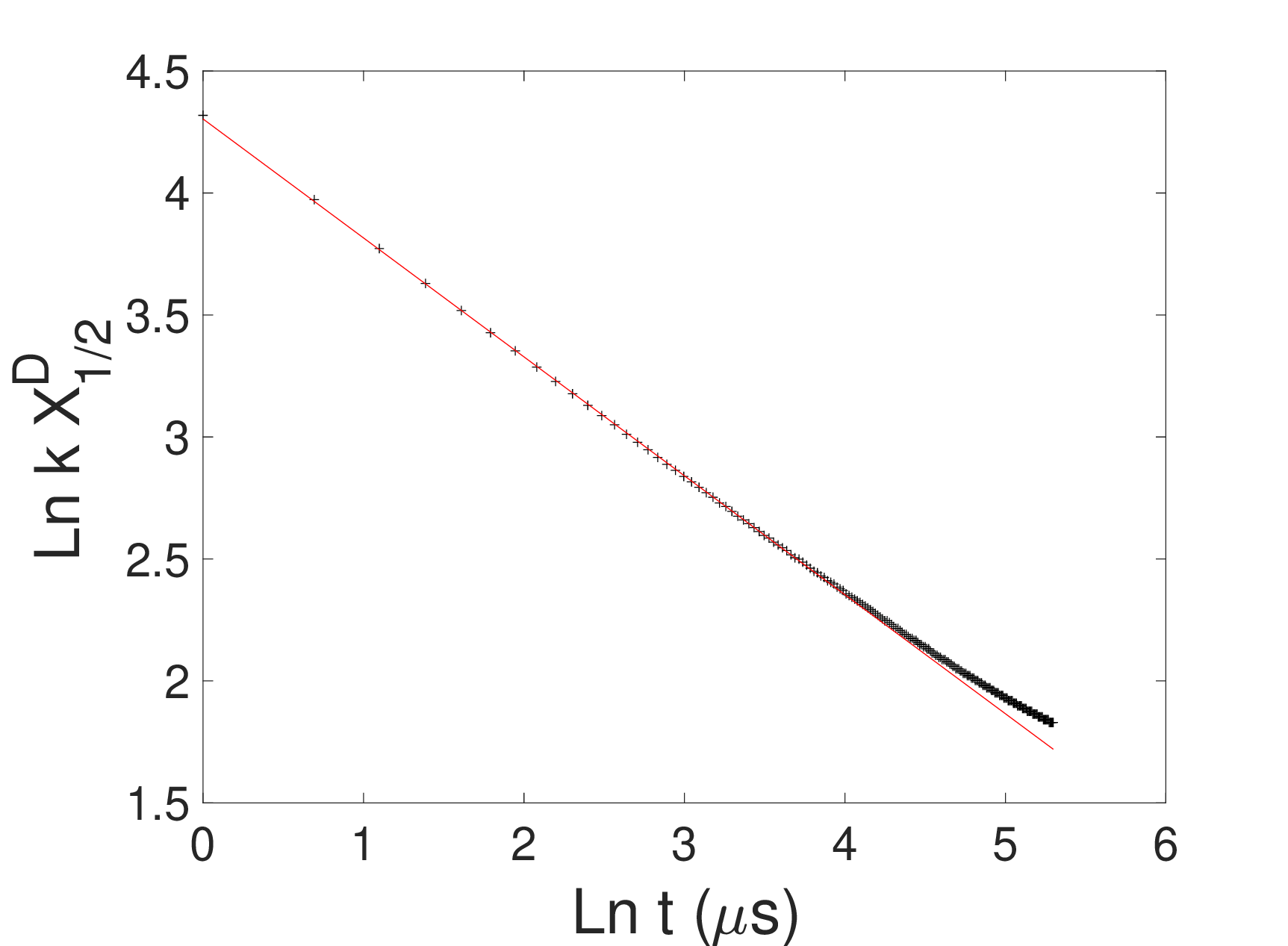}} 
 \subfigure[]{\includegraphics[width=.45 \textwidth]{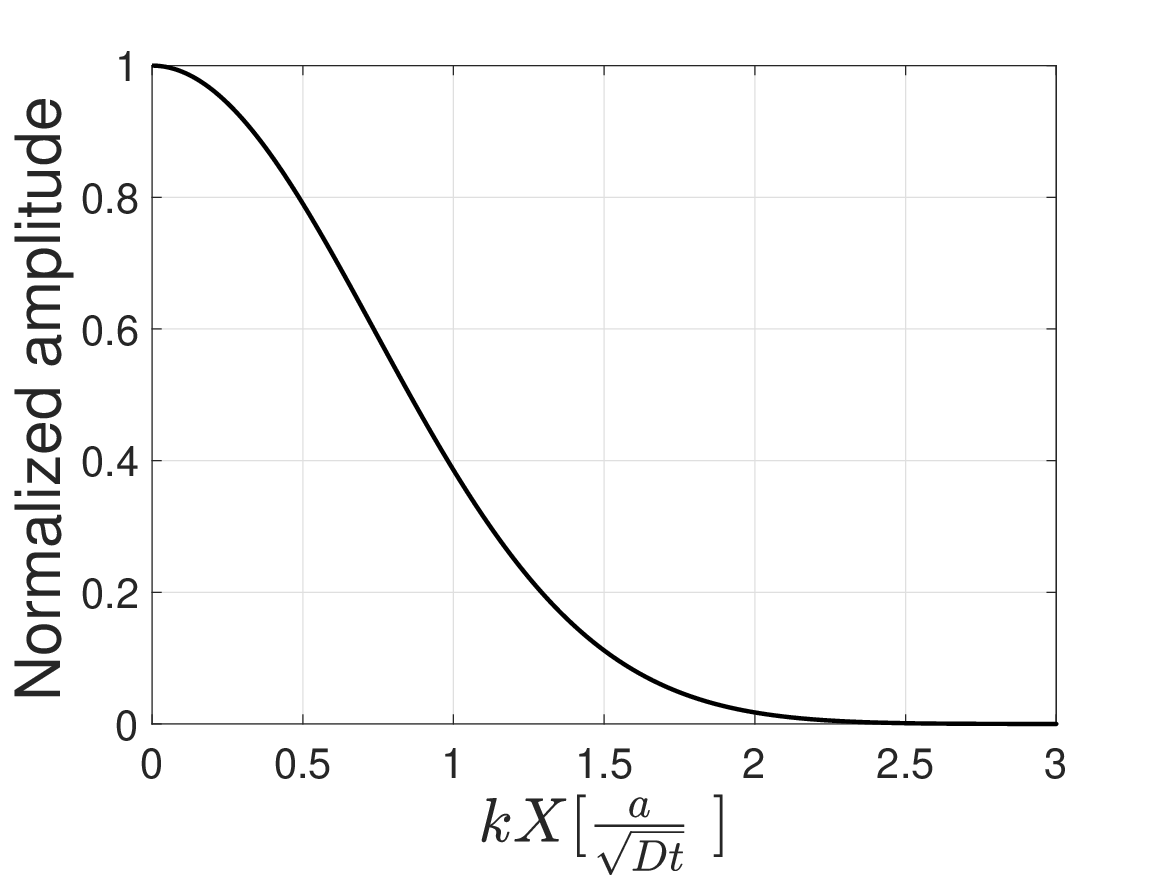}}
\caption{(a) The focal spot size is predicted to vary as the inverse square root of time (note the logarithmic scales on the axes.). (b) Dynamic focal spot in adimensional units.}
\label{fig:Dynamic_TR}
\end{figure}

In figure \ref{fig:Dynamic_TR}.a, the focal spot size is plotted as a function of time on a log-log scale. The $-1/2$ slope indicates an inverse square root dependence of the focal spot size and the y-intercept leads to the following formula for the half-width at half maximum (see figure \ref{fig:Dynamic_TR}.b)

\begin{equation}
X^D_{1/2} = \frac{0.82 a}{k \sqrt{Dt}} 
\label{eq:prediction}
\end{equation}

\bigskip

When the whole received signal is time-reversed and backpropagated, the resulting signal received at  $\textbf{r}_D$ at the focusing moment is given by:

\begin{equation*}
d^S(X)=F(\textbf{r}_D,\textbf{r}_S,t=0)
=\int_{-\infty}^{+\infty} G(\textbf{r}_S,\textbf{r}_T,\tau) G(\textbf{r}_D,\textbf{r}_T,\tau) d \tau
\end{equation*}

\bigskip

Time-reversal focusing in a disordered scattering medium is a self-averaging process \cite{derode2001random}, so that $d^S(X)$ is also given by:
\begin{multline*}
d^S(X)
=\int_{0}^{+\infty} \langle G(\textbf{r}_S,\textbf{r}_T,t) G(\textbf{r}_D,\textbf{r}_T,t)\rangle dt
\\
=\int_{0}^{+\infty}dt \frac{1}{(Dt)^{3/2}}\int_{0}^{\theta_{max}} d\theta \sin \theta J_0(kX \sin \theta) e^{-a^2 \tan^2 \theta /4Dt}  [e^{-z_T^2/4Dt}-e^{-(z_T + 2 z_0)^2/4Dt}]
\end{multline*}

The time integral involves the difference between two integrals of the form
\begin{equation*}
\int_{0}^{+\infty} \frac{e^{-b^2 /4Dt}}{(Dt)^{3/2}} dt  =\frac{2}{Db}\Gamma(1/2) = \frac{2 \sqrt{\pi}}{D b}
\end{equation*}
with $b^2=a^2 \tan^2 \theta +z_T^2$ or $b^2=a^2 \tan^2 \theta +(z_T+2z_0)^2$ 

Thus finally, 
\begin{equation*}
d^S(X) \propto \int_{0}^{\theta_{max}} d\theta \sin \theta J_0( kX \sin \theta)\left[ \frac{1}{\sqrt{a^2 \tan^2 \theta +z_T^2}}-\frac{1}{\sqrt{a^2 \tan^2 \theta +(z_T+2z_0)^2}} \right]
\end{equation*}
or
\begin{equation}
d^S(X) \propto \int_{0}^{\theta_{max}} d\theta \sin 2 \theta J_0(kX \sin \theta) \left[ \frac{1}{\sqrt{a^2 \sin^2 \theta +z_T^2 \cos^2 \theta}}-\frac{1}{\sqrt{a^2 \sin^2 \theta +\cos^2 \theta(z_T+2z_0)^2}} \right]
\label{eq:stat_foc_spot}
\end{equation}

As expected on the basis of the physical arguments invoked in the main text, the half-width at half-maximum of $d^S(X)$, $X^S_{1/2}$, deduced from Eq. \ref{eq:stat_foc_spot} is found to vary linearly against $1/z_T$ as soon as $a$ becomes larger than $z_T$ (see figure \ref{fig:Stationary_TR_Depth}.a). The existence of a non-zero value of $X^S_{1/2}$ for $z_T \gg a$  can be interpreted as a consequence of the diffraction limit. We have verified that the slope and y-intercept of $X^S_{1/2}$ against $a/z_T$ hardly depends on the choice for the extrapolation length $z_0$. In contrast, the finite source aperture and the finite duration of the TRW have a strong influence on both the y-intercept and the slope (see figure \ref{fig:Stationary_TR_Depth}.b). The y-intercept is found to depend on $ \theta_{max}$, which is not surprising given the physical argument that the minimum focal spot size cannot be smaller than the source size. The larger the source size, the smaller $ \theta_{max}$ and the larger the minimum focal spot. As to the slope, it is found to highly depend on the time window size.

\begin{figure}[H]
\begin{center}
 \subfigure[]{\includegraphics[width=.45 \textwidth]{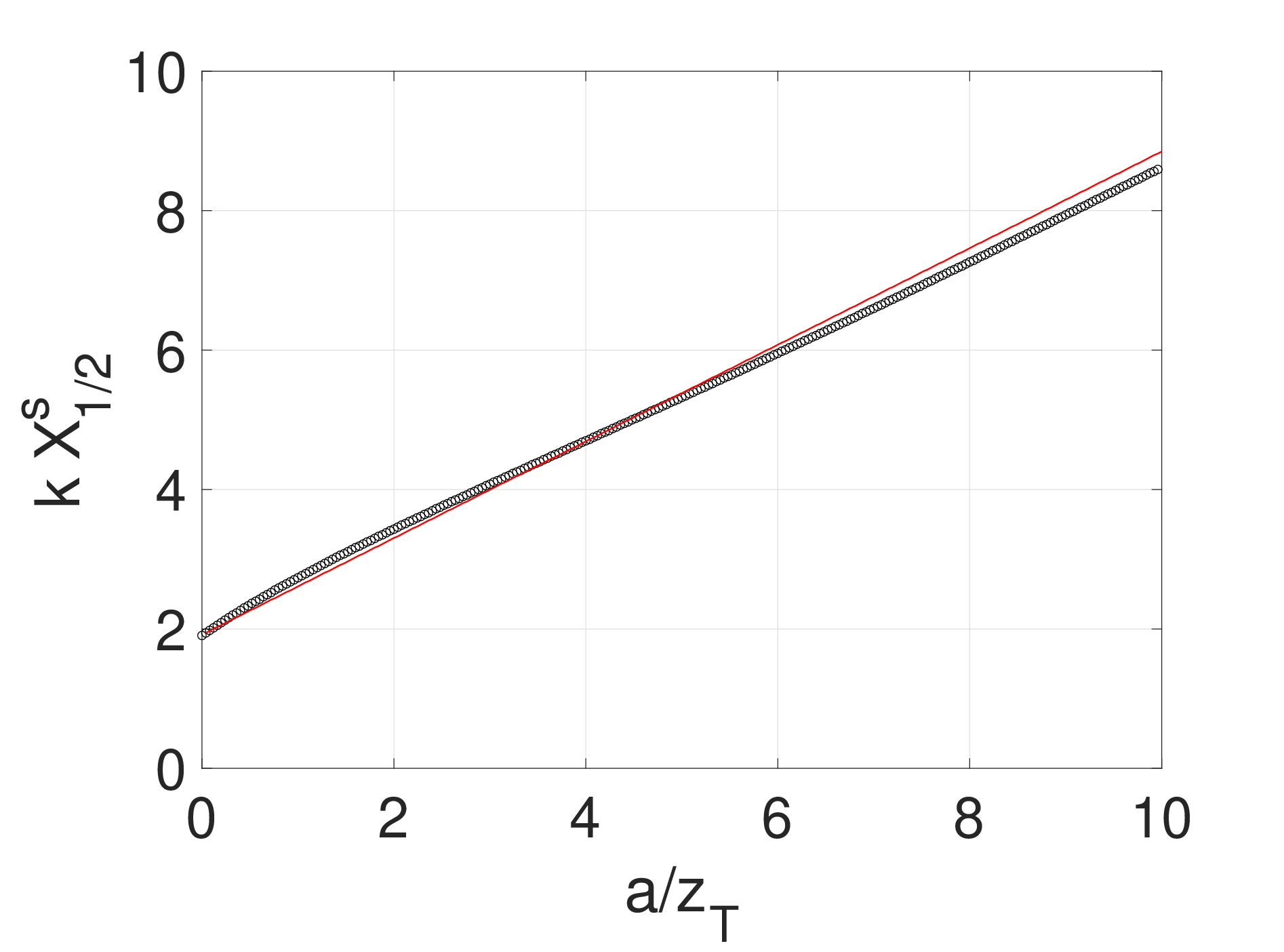}} 
  \subfigure[]{\includegraphics[width=.45 \textwidth]{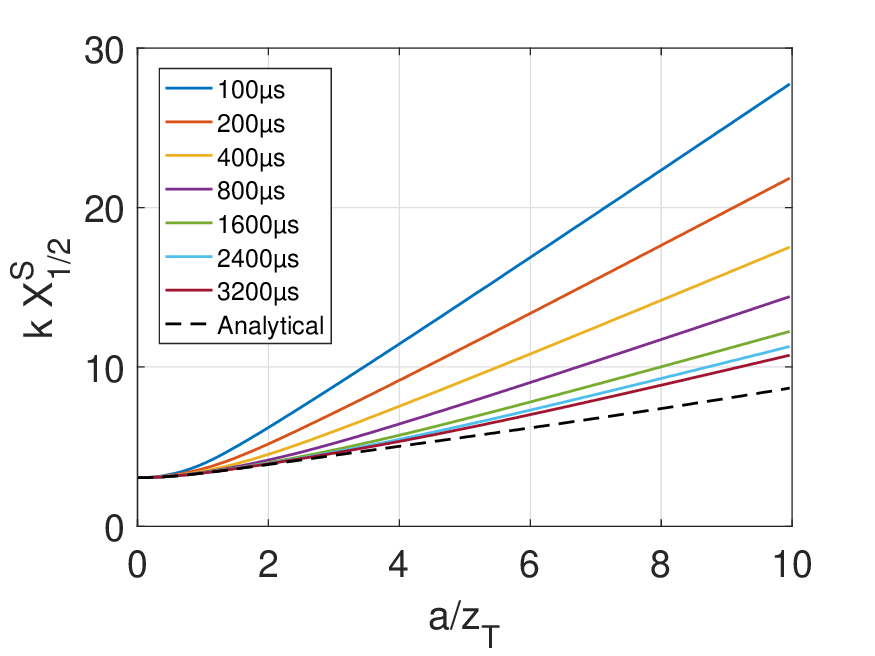}} 
\caption{Evolution of the stationary focal spot size as a function of inverse depth. (a) The black circles correspond to spot sizes deduced from Eq. 4 with $\theta = \pi / 2$. In red is plotted the prediction based on the Akkermans and Maynard formula for CBS  \cite{akkermans1986coherent}, and generalized in \cite{van2003green} for a TRM placed inside a scattering medium. It was shifted along the y-axis to make y-intercepts coincide. The two predictions are in quite good agreement. The y-intercept provides a minimum focal spot size of $\lambda / \pi$ corresponding to the diffraction limit. (b) Focal spot sizes obtained from Eq. \ref{eq:dyn} (with $\theta_{max}$ set to $\pi / 4$ to take the source directivity into account) numerically integrated over different time-window lengths. The curve obtained for $400 \mu s$, i.e., the size of the TRW used in our stationary time reversal experiment, is compared with the experimental data in figure 4 of the main text.}
\label{fig:Stationary_TR_Depth}
\end{center}
\end{figure}

\bibliography{bibliography}

\begin{thebibliography}{24}%
\makeatletter
\providecommand \@ifxundefined [1]{%
 \@ifx{#1\undefined}
}%
\providecommand \@ifnum [1]{%
 \ifnum #1\expandafter \@firstoftwo
 \else \expandafter \@secondoftwo
 \fi
}%
\providecommand \@ifx [1]{%
 \ifx #1\expandafter \@firstoftwo
 \else \expandafter \@secondoftwo
 \fi
}%
\providecommand \natexlab [1]{#1}%
\providecommand \enquote  [1]{``#1''}%
\providecommand \bibnamefont  [1]{#1}%
\providecommand \bibfnamefont [1]{#1}%
\providecommand \citenamefont [1]{#1}%
\providecommand \href@noop [0]{\@secondoftwo}%
\providecommand \href [0]{\begingroup \@sanitize@url \@href}%
\providecommand \@href[1]{\@@startlink{#1}\@@href}%
\providecommand \@@href[1]{\endgroup#1\@@endlink}%
\providecommand \@sanitize@url [0]{\catcode `\\12\catcode `\$12\catcode
  `\&12\catcode `\#12\catcode `\^12\catcode `\_12\catcode `\%12\relax}%
\providecommand \@@startlink[1]{}%
\providecommand \@@endlink[0]{}%
\providecommand \url  [0]{\begingroup\@sanitize@url \@url }%
\providecommand \@url [1]{\endgroup\@href {#1}{\urlprefix }}%
\providecommand \urlprefix  [0]{URL }%
\providecommand \Eprint [0]{\href }%
\providecommand \doibase [0]{https://doi.org/}%
\providecommand \selectlanguage [0]{\@gobble}%
\providecommand \bibinfo  [0]{\@secondoftwo}%
\providecommand \bibfield  [0]{\@secondoftwo}%
\providecommand \translation [1]{[#1]}%
\providecommand \BibitemOpen [0]{}%
\providecommand \bibitemStop [0]{}%
\providecommand \bibitemNoStop [0]{.\EOS\space}%
\providecommand \EOS [0]{\spacefactor3000\relax}%
\providecommand \BibitemShut  [1]{\csname bibitem#1\endcsname}%
\let\auto@bib@innerbib\@empty
\bibitem [{\citenamefont {Sheng}(2006)}]{sheng2006introduction}%
  \BibitemOpen
  \bibfield  {author} {\bibinfo {author} {\bibfnamefont {P.}~\bibnamefont
  {Sheng}},\ }\href@noop {} {\emph {\bibinfo {title} {Introduction to wave
  scattering, localization and mesoscopic phenomena}}},\ Vol.~\bibinfo {volume}
  {88}\ (\bibinfo  {publisher} {Springer Science \& Business Media},\ \bibinfo
  {year} {2006})\BibitemShut {NoStop}%
\bibitem [{\citenamefont {Page}\ \emph {et~al.}(1995)\citenamefont {Page},
  \citenamefont {Schriemer}, \citenamefont {Bailey},\ and\ \citenamefont
  {Weitz}}]{page1995experimental}%
  \BibitemOpen
  \bibfield  {author} {\bibinfo {author} {\bibfnamefont {J.}~\bibnamefont
  {Page}}, \bibinfo {author} {\bibfnamefont {H.}~\bibnamefont {Schriemer}},
  \bibinfo {author} {\bibfnamefont {A.}~\bibnamefont {Bailey}},\ and\ \bibinfo
  {author} {\bibfnamefont {D.}~\bibnamefont {Weitz}},\ }\bibfield  {title}
  {\bibinfo {title} {Experimental test of the diffusion approximation for
  multiply scattered sound},\ }\href@noop {} {\bibfield  {journal} {\bibinfo
  {journal} {Physical Review E}\ }\textbf {\bibinfo {volume} {52}},\ \bibinfo
  {pages} {3106} (\bibinfo {year} {1995})}\BibitemShut {NoStop}%
\bibitem [{\citenamefont {Tourin}\ \emph {et~al.}(2000)\citenamefont {Tourin},
  \citenamefont {Derode}, \citenamefont {Peyre},\ and\ \citenamefont
  {Fink}}]{tourin2000transport}%
  \BibitemOpen
  \bibfield  {author} {\bibinfo {author} {\bibfnamefont {A.}~\bibnamefont
  {Tourin}}, \bibinfo {author} {\bibfnamefont {A.}~\bibnamefont {Derode}},
  \bibinfo {author} {\bibfnamefont {A.}~\bibnamefont {Peyre}},\ and\ \bibinfo
  {author} {\bibfnamefont {M.}~\bibnamefont {Fink}},\ }\bibfield  {title}
  {\bibinfo {title} {Transport parameters for an ultrasonic pulsed wave
  propagating in a multiple scattering medium},\ }\href@noop {} {\bibfield
  {journal} {\bibinfo  {journal} {The Journal of the Acoustical Society of
  America}\ }\textbf {\bibinfo {volume} {108}},\ \bibinfo {pages} {503}
  (\bibinfo {year} {2000})}\BibitemShut {NoStop}%
\bibitem [{\citenamefont {Derode}\ \emph {et~al.}(1995)\citenamefont {Derode},
  \citenamefont {Roux},\ and\ \citenamefont {Fink}}]{derode1995robust}%
  \BibitemOpen
  \bibfield  {author} {\bibinfo {author} {\bibfnamefont {A.}~\bibnamefont
  {Derode}}, \bibinfo {author} {\bibfnamefont {P.}~\bibnamefont {Roux}},\ and\
  \bibinfo {author} {\bibfnamefont {M.}~\bibnamefont {Fink}},\ }\bibfield
  {title} {\bibinfo {title} {Robust acoustic time reversal with high-order
  multiple scattering},\ }\href@noop {} {\bibfield  {journal} {\bibinfo
  {journal} {Physical review letters}\ }\textbf {\bibinfo {volume} {75}},\
  \bibinfo {pages} {4206} (\bibinfo {year} {1995})}\BibitemShut {NoStop}%
\bibitem [{\citenamefont {Tourin}\ \emph {et~al.}(1999)\citenamefont {Tourin},
  \citenamefont {Derode},\ and\ \citenamefont {Fink}}]{tourin1999dynamic}%
  \BibitemOpen
  \bibfield  {author} {\bibinfo {author} {\bibfnamefont {A.}~\bibnamefont
  {Tourin}}, \bibinfo {author} {\bibfnamefont {A.}~\bibnamefont {Derode}},\
  and\ \bibinfo {author} {\bibfnamefont {M.}~\bibnamefont {Fink}},\ }\bibfield
  {title} {\bibinfo {title} {Dynamic time reversal of randomly backscattered
  acoustic waves},\ }\href@noop {} {\bibfield  {journal} {\bibinfo  {journal}
  {EPL (Europhysics Letters)}\ }\textbf {\bibinfo {volume} {47}},\ \bibinfo
  {pages} {175} (\bibinfo {year} {1999})}\BibitemShut {NoStop}%
\bibitem [{\citenamefont {Tourin}\ \emph {et~al.}(2006)\citenamefont {Tourin},
  \citenamefont {Van Der~Biest},\ and\ \citenamefont
  {Fink}}]{tourin2006phononic}%
  \BibitemOpen
  \bibfield  {author} {\bibinfo {author} {\bibfnamefont {A.}~\bibnamefont
  {Tourin}}, \bibinfo {author} {\bibfnamefont {F.}~\bibnamefont {Van
  Der~Biest}},\ and\ \bibinfo {author} {\bibfnamefont {M.}~\bibnamefont
  {Fink}},\ }\bibfield  {title} {\bibinfo {title} {Time reversal of ultrasound
  through a phononic crystal},\ }\href
  {https://doi.org/10.1103/PhysRevLett.96.104301} {\bibfield  {journal}
  {\bibinfo  {journal} {Phys. Rev. Lett.}\ }\textbf {\bibinfo {volume} {96}},\
  \bibinfo {pages} {104301} (\bibinfo {year} {2006})}\BibitemShut {NoStop}%
\bibitem [{\citenamefont {Van~Tiggelen}(2003)}]{van2003green}%
  \BibitemOpen
  \bibfield  {author} {\bibinfo {author} {\bibfnamefont {B.}~\bibnamefont
  {Van~Tiggelen}},\ }\bibfield  {title} {\bibinfo {title} {Green function
  retrieval and time reversal in a disordered world},\ }\href@noop {}
  {\bibfield  {journal} {\bibinfo  {journal} {Physical review letters}\
  }\textbf {\bibinfo {volume} {91}},\ \bibinfo {pages} {243904} (\bibinfo
  {year} {2003})}\BibitemShut {NoStop}%
\bibitem [{\citenamefont {Schriemer}\ \emph {et~al.}(1997)\citenamefont
  {Schriemer}, \citenamefont {Cowan}, \citenamefont {Page}, \citenamefont
  {Sheng}, \citenamefont {Liu},\ and\ \citenamefont
  {Weitz}}]{schriemer1997energy}%
  \BibitemOpen
  \bibfield  {author} {\bibinfo {author} {\bibfnamefont {H.~P.}\ \bibnamefont
  {Schriemer}}, \bibinfo {author} {\bibfnamefont {M.~L.}\ \bibnamefont
  {Cowan}}, \bibinfo {author} {\bibfnamefont {J.~H.}\ \bibnamefont {Page}},
  \bibinfo {author} {\bibfnamefont {P.}~\bibnamefont {Sheng}}, \bibinfo
  {author} {\bibfnamefont {Z.}~\bibnamefont {Liu}},\ and\ \bibinfo {author}
  {\bibfnamefont {D.~A.}\ \bibnamefont {Weitz}},\ }\bibfield  {title} {\bibinfo
  {title} {Energy velocity of diffusing waves in strongly scattering media},\
  }\href@noop {} {\bibfield  {journal} {\bibinfo  {journal} {Physical Review
  Letters}\ }\textbf {\bibinfo {volume} {79}},\ \bibinfo {pages} {3166}
  (\bibinfo {year} {1997})}\BibitemShut {NoStop}%
\bibitem [{\citenamefont {Derode}\ \emph {et~al.}(2001)\citenamefont {Derode},
  \citenamefont {Tourin},\ and\ \citenamefont {Fink}}]{derode2001random}%
  \BibitemOpen
  \bibfield  {author} {\bibinfo {author} {\bibfnamefont {A.}~\bibnamefont
  {Derode}}, \bibinfo {author} {\bibfnamefont {A.}~\bibnamefont {Tourin}},\
  and\ \bibinfo {author} {\bibfnamefont {M.}~\bibnamefont {Fink}},\ }\bibfield
  {title} {\bibinfo {title} {Random multiple scattering of ultrasound. ii. is
  time reversal a self-averaging process?},\ }\href@noop {} {\bibfield
  {journal} {\bibinfo  {journal} {Physical Review E}\ }\textbf {\bibinfo
  {volume} {64}},\ \bibinfo {pages} {036606} (\bibinfo {year}
  {2001})}\BibitemShut {NoStop}%
\bibitem [{\citenamefont {Akkermans}\ \emph {et~al.}(1986)\citenamefont
  {Akkermans}, \citenamefont {Wolf},\ and\ \citenamefont
  {Maynard}}]{akkermans1986coherent}%
  \BibitemOpen
  \bibfield  {author} {\bibinfo {author} {\bibfnamefont {E.}~\bibnamefont
  {Akkermans}}, \bibinfo {author} {\bibfnamefont {P.}~\bibnamefont {Wolf}},\
  and\ \bibinfo {author} {\bibfnamefont {R.}~\bibnamefont {Maynard}},\
  }\bibfield  {title} {\bibinfo {title} {Coherent backscattering of light by
  disordered media: Analysis of the peak line shape},\ }\href@noop {}
  {\bibfield  {journal} {\bibinfo  {journal} {Physical review letters}\
  }\textbf {\bibinfo {volume} {56}},\ \bibinfo {pages} {1471} (\bibinfo {year}
  {1986})}\BibitemShut {NoStop}%
\bibitem [{\citenamefont {Goodman}(2015)}]{goodman2015statistical}%
  \BibitemOpen
  \bibfield  {author} {\bibinfo {author} {\bibfnamefont {J.~W.}\ \bibnamefont
  {Goodman}},\ }\href@noop {} {\emph {\bibinfo {title} {Statistical optics}}}\
  (\bibinfo  {publisher} {John Wiley \& Sons},\ \bibinfo {year}
  {2015})\BibitemShut {NoStop}%
\bibitem [{\citenamefont {Tsang}\ and\ \citenamefont
  {Ishimaru}(1984)}]{Tsang:84}%
  \BibitemOpen
  \bibfield  {author} {\bibinfo {author} {\bibfnamefont {L.}~\bibnamefont
  {Tsang}}\ and\ \bibinfo {author} {\bibfnamefont {A.}~\bibnamefont
  {Ishimaru}},\ }\bibfield  {title} {\bibinfo {title} {Backscattering
  enhancement of random discrete scatterers},\ }\href
  {https://doi.org/10.1364/JOSAA.1.000836} {\bibfield  {journal} {\bibinfo
  {journal} {J. Opt. Soc. Am. A}\ }\textbf {\bibinfo {volume} {1}},\ \bibinfo
  {pages} {836} (\bibinfo {year} {1984})}\BibitemShut {NoStop}%
\bibitem [{\citenamefont {Wolf}\ and\ \citenamefont
  {Maret}(1985)}]{wolf1985weak}%
  \BibitemOpen
  \bibfield  {author} {\bibinfo {author} {\bibfnamefont {P.-E.}\ \bibnamefont
  {Wolf}}\ and\ \bibinfo {author} {\bibfnamefont {G.}~\bibnamefont {Maret}},\
  }\bibfield  {title} {\bibinfo {title} {Weak localization and coherent
  backscattering of photons in disordered media},\ }\href@noop {} {\bibfield
  {journal} {\bibinfo  {journal} {Physical review letters}\ }\textbf {\bibinfo
  {volume} {55}},\ \bibinfo {pages} {2696} (\bibinfo {year}
  {1985})}\BibitemShut {NoStop}%
\bibitem [{\citenamefont {Van~Albada}\ and\ \citenamefont
  {Lagendijk}(1985)}]{van1985observation}%
  \BibitemOpen
  \bibfield  {author} {\bibinfo {author} {\bibfnamefont {M.~P.}\ \bibnamefont
  {Van~Albada}}\ and\ \bibinfo {author} {\bibfnamefont {A.}~\bibnamefont
  {Lagendijk}},\ }\bibfield  {title} {\bibinfo {title} {Observation of weak
  localization of light in a random medium},\ }\href@noop {} {\bibfield
  {journal} {\bibinfo  {journal} {Physical review letters}\ }\textbf {\bibinfo
  {volume} {55}},\ \bibinfo {pages} {2692} (\bibinfo {year}
  {1985})}\BibitemShut {NoStop}%
\bibitem [{\citenamefont {Bayer}\ and\ \citenamefont
  {Niederdr{\"a}nk}(1993)}]{bayer1993weak}%
  \BibitemOpen
  \bibfield  {author} {\bibinfo {author} {\bibfnamefont {G.}~\bibnamefont
  {Bayer}}\ and\ \bibinfo {author} {\bibfnamefont {T.}~\bibnamefont
  {Niederdr{\"a}nk}},\ }\bibfield  {title} {\bibinfo {title} {Weak localization
  of acoustic waves in strongly scattering media},\ }\href@noop {} {\bibfield
  {journal} {\bibinfo  {journal} {Physical review letters}\ }\textbf {\bibinfo
  {volume} {70}},\ \bibinfo {pages} {3884} (\bibinfo {year}
  {1993})}\BibitemShut {NoStop}%
\bibitem [{\citenamefont {Tourin}\ \emph {et~al.}(1997)\citenamefont {Tourin},
  \citenamefont {Derode}, \citenamefont {Roux}, \citenamefont {Van~Tiggelen},\
  and\ \citenamefont {Fink}}]{tourin1997time}%
  \BibitemOpen
  \bibfield  {author} {\bibinfo {author} {\bibfnamefont {A.}~\bibnamefont
  {Tourin}}, \bibinfo {author} {\bibfnamefont {A.}~\bibnamefont {Derode}},
  \bibinfo {author} {\bibfnamefont {P.}~\bibnamefont {Roux}}, \bibinfo {author}
  {\bibfnamefont {B.~A.}\ \bibnamefont {Van~Tiggelen}},\ and\ \bibinfo {author}
  {\bibfnamefont {M.}~\bibnamefont {Fink}},\ }\bibfield  {title} {\bibinfo
  {title} {Time-dependent coherent backscattering of acoustic waves},\
  }\href@noop {} {\bibfield  {journal} {\bibinfo  {journal} {Physical review
  letters}\ }\textbf {\bibinfo {volume} {79}},\ \bibinfo {pages} {3637}
  (\bibinfo {year} {1997})}\BibitemShut {NoStop}%
\bibitem [{\citenamefont {De~Rosny}\ \emph {et~al.}(2004)\citenamefont
  {De~Rosny}, \citenamefont {Tourin}, \citenamefont {Derode}, \citenamefont
  {Van~Tiggelen},\ and\ \citenamefont {Fink}}]{de2004relation}%
  \BibitemOpen
  \bibfield  {author} {\bibinfo {author} {\bibfnamefont {J.}~\bibnamefont
  {De~Rosny}}, \bibinfo {author} {\bibfnamefont {A.}~\bibnamefont {Tourin}},
  \bibinfo {author} {\bibfnamefont {A.}~\bibnamefont {Derode}}, \bibinfo
  {author} {\bibfnamefont {B.}~\bibnamefont {Van~Tiggelen}},\ and\ \bibinfo
  {author} {\bibfnamefont {M.}~\bibnamefont {Fink}},\ }\bibfield  {title}
  {\bibinfo {title} {Relation between time reversal focusing and coherent
  backscattering in multiple scattering media: A diagrammatic approach},\
  }\href@noop {} {\bibfield  {journal} {\bibinfo  {journal} {Physical Review
  E}\ }\textbf {\bibinfo {volume} {70}},\ \bibinfo {pages} {046601} (\bibinfo
  {year} {2004})}\BibitemShut {NoStop}%
\bibitem [{\citenamefont {Papanicolaou}\ \emph {et~al.}(2004)\citenamefont
  {Papanicolaou}, \citenamefont {Solna},\ and\ \citenamefont
  {Ryzhik}}]{Papanicolaou2002}%
  \BibitemOpen
  \bibfield  {author} {\bibinfo {author} {\bibfnamefont {G.}~\bibnamefont
  {Papanicolaou}}, \bibinfo {author} {\bibfnamefont {K.}~\bibnamefont
  {Solna}},\ and\ \bibinfo {author} {\bibfnamefont {L.}~\bibnamefont
  {Ryzhik}},\ }\bibfield  {title} {\bibinfo {title} {Statistical stability in
  time reversal},\ }\href@noop {} {\bibfield  {journal} {\bibinfo  {journal}
  {SIAM Journal on Applied Mathematics}\ }\textbf {\bibinfo {volume} {64}},\
  \bibinfo {pages} {1133} (\bibinfo {year} {2004})}\BibitemShut {NoStop}%
\bibitem [{\citenamefont {Cobus}\ \emph {et~al.}(2017)\citenamefont {Cobus},
  \citenamefont {Van~Tiggelen}, \citenamefont {Derode},\ and\ \citenamefont
  {Page}}]{cobus2017dynamic}%
  \BibitemOpen
  \bibfield  {author} {\bibinfo {author} {\bibfnamefont {L.}~\bibnamefont
  {Cobus}}, \bibinfo {author} {\bibfnamefont {B.}~\bibnamefont {Van~Tiggelen}},
  \bibinfo {author} {\bibfnamefont {A.}~\bibnamefont {Derode}},\ and\ \bibinfo
  {author} {\bibfnamefont {J.}~\bibnamefont {Page}},\ }\bibfield  {title}
  {\bibinfo {title} {Dynamic coherent backscattering of ultrasound in
  three-dimensional strongly-scattering media},\ }\href@noop {} {\bibfield
  {journal} {\bibinfo  {journal} {The European Physical Journal Special
  Topics}\ }\textbf {\bibinfo {volume} {226}},\ \bibinfo {pages} {1549}
  (\bibinfo {year} {2017})}\BibitemShut {NoStop}%
\bibitem [{\citenamefont {Backlund}\ \emph {et~al.}(2018)\citenamefont
  {Backlund}, \citenamefont {Shechtman},\ and\ \citenamefont
  {Walsworth}}]{PhysRevLett.121.023904}%
  \BibitemOpen
  \bibfield  {author} {\bibinfo {author} {\bibfnamefont {M.~P.}\ \bibnamefont
  {Backlund}}, \bibinfo {author} {\bibfnamefont {Y.}~\bibnamefont
  {Shechtman}},\ and\ \bibinfo {author} {\bibfnamefont {R.~L.}\ \bibnamefont
  {Walsworth}},\ }\bibfield  {title} {\bibinfo {title} {Fundamental precision
  bounds for three-dimensional optical localization microscopy with poisson
  statistics},\ }\href {https://doi.org/10.1103/PhysRevLett.121.023904}
  {\bibfield  {journal} {\bibinfo  {journal} {Phys. Rev. Lett.}\ }\textbf
  {\bibinfo {volume} {121}},\ \bibinfo {pages} {023904} (\bibinfo {year}
  {2018})}\BibitemShut {NoStop}%
\bibitem [{\citenamefont {Ulrich}\ \emph {et~al.}(2007)\citenamefont {Ulrich},
  \citenamefont {Johnson},\ and\ \citenamefont
  {Guyer}}]{PhysRevLett.98.104301}%
  \BibitemOpen
  \bibfield  {author} {\bibinfo {author} {\bibfnamefont {T.~J.}\ \bibnamefont
  {Ulrich}}, \bibinfo {author} {\bibfnamefont {P.~A.}\ \bibnamefont
  {Johnson}},\ and\ \bibinfo {author} {\bibfnamefont {R.~A.}\ \bibnamefont
  {Guyer}},\ }\bibfield  {title} {\bibinfo {title} {Interaction dynamics of
  elastic waves with a complex nonlinear scatterer through the use of a time
  reversal mirror},\ }\href {https://doi.org/10.1103/PhysRevLett.98.104301}
  {\bibfield  {journal} {\bibinfo  {journal} {Phys. Rev. Lett.}\ }\textbf
  {\bibinfo {volume} {98}},\ \bibinfo {pages} {104301} (\bibinfo {year}
  {2007})}\BibitemShut {NoStop}%
\bibitem [{\citenamefont {Mostajabi}\ \emph {et~al.}(2019)\citenamefont
  {Mostajabi}, \citenamefont {Karami}, \citenamefont {Azadifar}, \citenamefont
  {Ghasemi}, \citenamefont {Rubinstein},\ and\ \citenamefont
  {Rachidi}}]{Scientific_Report}%
  \BibitemOpen
  \bibfield  {author} {\bibinfo {author} {\bibfnamefont {A.}~\bibnamefont
  {Mostajabi}}, \bibinfo {author} {\bibfnamefont {H.}~\bibnamefont {Karami}},
  \bibinfo {author} {\bibfnamefont {M.}~\bibnamefont {Azadifar}}, \bibinfo
  {author} {\bibfnamefont {A.}~\bibnamefont {Ghasemi}}, \bibinfo {author}
  {\bibfnamefont {M.}~\bibnamefont {Rubinstein}},\ and\ \bibinfo {author}
  {\bibfnamefont {F.}~\bibnamefont {Rachidi}},\ }\bibfield  {title} {\bibinfo
  {title} {Single-sensor source localization using electromagnetic time
  reversal and deep transfer learning: Application to lightning},\ }\href
  {https://doi.org/10.1038/s41598-019-53934-4} {\bibfield  {journal} {\bibinfo
  {journal} {Scientific Report}\ }\textbf {\bibinfo {volume} {9}},\ \bibinfo
  {pages} {17372} (\bibinfo {year} {2019})}\BibitemShut {NoStop}%
\bibitem [{\citenamefont {Lagendijk}\ and\ \citenamefont {{van
  Tiggelen}}(1996)}]{LAGENDIJK1996143}%
  \BibitemOpen
  \bibfield  {author} {\bibinfo {author} {\bibfnamefont {A.}~\bibnamefont
  {Lagendijk}}\ and\ \bibinfo {author} {\bibfnamefont {B.~A.}\ \bibnamefont
  {{van Tiggelen}}},\ }\bibfield  {title} {\bibinfo {title} {Resonant multiple
  scattering of light},\ }\href
  {https://doi.org/https://doi.org/10.1016/0370-1573(95)00065-8} {\bibfield
  {journal} {\bibinfo  {journal} {Physics Reports}\ }\textbf {\bibinfo {volume}
  {270}},\ \bibinfo {pages} {143} (\bibinfo {year} {1996})}\BibitemShut
  {NoStop}%
\bibitem [{\citenamefont {Lagendijk}\ \emph {et~al.}(1989)\citenamefont
  {Lagendijk}, \citenamefont {Vreeker},\ and\ \citenamefont
  {De~Vries}}]{lagendijk1989influence}%
  \BibitemOpen
  \bibfield  {author} {\bibinfo {author} {\bibfnamefont {A.}~\bibnamefont
  {Lagendijk}}, \bibinfo {author} {\bibfnamefont {R.}~\bibnamefont {Vreeker}},\
  and\ \bibinfo {author} {\bibfnamefont {P.}~\bibnamefont {De~Vries}},\
  }\bibfield  {title} {\bibinfo {title} {Influence of internal reflection on
  diffusive transport in strongly scattering media},\ }\href@noop {} {\bibfield
   {journal} {\bibinfo  {journal} {Physics Letters A}\ }\textbf {\bibinfo
  {volume} {136}},\ \bibinfo {pages} {81} (\bibinfo {year} {1989})}\BibitemShut
  {NoStop}%
\end{thebibliography}%

\end{document}